\pdfoutput=1
\documentclass[prd,nofootinbib,preprintnumbers,twocolumn,
showpacs,floatfix]{revtex4-1}

 \usepackage{auncial}
\usepackage{scrextend}
\usepackage{relsize}
\usepackage{amsmath}
\usepackage{amssymb}
\usepackage{epsfig}
\usepackage{graphicx}
\usepackage{hyperref}
\usepackage{dcolumn}   
\usepackage{tabu}
\usepackage{boldline}
\usepackage{slashed}
\usepackage{multirow}
\usepackage{color}
\usepackage[normal]{subfigure}
\usepackage{rotating}
\usepackage[margin=0.9in,a4paper]{geometry}
\usepackage[table]{xcolor}
\usepackage{enumitem}
\usepackage[utf8]{inputenc}
\usepackage{colortbl}
\usepackage{array,multirow}
\definecolor{nicered}{rgb}{0.7,0.1,0.1}
\definecolor{nicegreen}{rgb}{0.1,0.5,0.1}
\definecolor{red}{rgb}{1.0, 0, 0}

\def\eq#1{{Eq.~(\ref{#1})}}

\def\gsim{\raise0.3ex\hbox{$\;>$\kern-0.75em\raise-1.1ex\hbox{$\sim\;$}}}
\def\lsim{\raise0.3ex\hbox{$\;<$\kern-0.75em\raise-1.1ex\hbox{$\sim\;$}}}

\def\mb[#1]{\mathbf{#1}}

\renewcommand{\bar}{\overline}

\definecolor{LightCyan}{rgb}{0.88,1,1}
\definecolor{piggypink}{rgb}{0.99, 0.87, 0.9}
\definecolor{applegreen}{rgb}{0.55, 0.71, 0.0}
\definecolor{darkpastelgreen}{rgb}{0.01, 0.75, 0.24}
\definecolor{green-yellow}{rgb}{0.68, 1.0, 0.18}

\newcommand{\beq}{\begin{equation}}
\newcommand{\eeq}{\end{equation}}
\newcommand{\beqa}{\begin{eqnarray}}
\newcommand{\eeqa}{\end{eqnarray}}

\newcommand{\veps}{\varepsilon}
\newcommand{\bveps}{\boldsymbol{\veps}}

\newcommand{\ca}[1]{{\color{blue} #1}}

\newcommand{\AddrMainz}{%
PRISMA Cluster of Excellence \& Mainz Institute for Theoretical Physics, Johannes Gutenberg-Universit\"at Mainz, 55099 Mainz, Germany
}

\begin{document}


\title{Audible Axions}

\author{Camila S. Machado}
\email{camachad@uni-mainz.de}
\author{Wolfram Ratzinger}
\email{wratzing@students.uni-mainz.de}
\author{Pedro Schwaller}
\email{pedro.schwaller@uni-mainz.de}
\author{Ben A. Stefanek\vspace{6pt}
}
\email{bstefan@uni-mainz.de}
\affiliation{\normalsize\it \AddrMainz}

\begin{abstract}
  \noindent

Conventional approaches to probing axions and axion-like particles (ALPs) typically rely on a coupling to photons. However, if this coupling is extremely weak, ALPs become invisible and are effectively decoupled from the Standard Model. Here we show that such invisible axions, which are viable candidates for dark matter, can produce a stochastic gravitational wave background in the early universe. This signal is generated in models where the invisible axion couples to a dark gauge boson that experiences a tachyonic instability when the axion begins to oscillate. Incidentally, the same mechanism also widens the viable parameter space for axion dark matter. Quantum fluctuations amplified by the exponentially growing gauge boson modes source chiral gravitational waves. For axion decay constants $f\gtrsim 10^{17}$~GeV, this signal is detectable by either pulsar timing arrays or space/ground-based gravitational wave detectors for a broad range of axion masses, thus providing a new window to probe invisible axion models.

\end{abstract}

\maketitle

\section{Introduction}
Axions, or more generally, axion like particles (ALPs) appear in many extensions of the Standard Model (SM). This includes solutions to the strong CP problem via the Peccei-Quinn mechanism~\cite{Peccei:1977ur,Peccei:1977hh}, string theory~\cite{Svrcek:2006yi,Arvanitaki:2009fg}, natural models of inflation~\cite{Freese:1990rb}, dark matter (DM)~\cite{Abbott:1982af,Dine:1982ah,Preskill:1982cy}, the relaxion mechanism for solving the hierarchy problem~\cite{Graham:2015cka}, or just in general models where an approximate global symmetry is spontaneously broken at a high scale, resulting in a very light pseudo Nambu-Goldstone boson. The viable parameter space for ALP masses and couplings spans many orders of magnitude, which makes searching for them challenging, but also motivates new ideas and approaches to probe previously inaccessible regions. In particular, if the ALP is effectively decoupled from the SM, the only bound comes from black-hole superradiance~\cite{Cardoso:2018tly}.

Here we show that axions and ALPs may leave a trace of their presence in the early universe in the form of a stochastic gravitational wave (GW) background. This is possible if the axion couples to a dark photon. 
Initially, the axion is displaced from its minimum and Hubble friction prevents it from rolling until the Hubble rate drops below the axion mass. Once it begins to roll, the axion induces a tachyonic instability for one of the dark photon helicities, causing vacuum  fluctuations to grow exponentially. This induces time-dependent anisotropic stress in the energy-momentum tensor, which ultimately sources gravitational waves. In the process, a large fraction of the energy density stored in the axion field is converted into radiation in the form of dark photons and gravitational waves. 

Subsequently, a period of oscillation occurs where the axion undergoes parametric resonance, further suppressing the amount of energy stored in the axion field. Observable gravitational wave signals require that the energy density stored in the axion field at the time of GW emission is large, so the combination of the tachyonic instability and the subsequent parametric resonance is necessary to prevent overabundant axion dark matter unless the axion is heavy enough to decay. This mechanism for efficient depletion of axion DM abundance via exponential production of dark photons was first pointed out in Ref.~\cite{Agrawal:2017eqm}, where it was used to increase the QCD axion decay constant without tuning the initial conditions. It was also noted that the SM photon cannot play the role of the gauge boson because its fast thermalization rate would destroy the conditions required for exponential particle production. 

The tachyonic instability of rolling ALPs has been previously  exploited in a variety of contexts, e.g. inflationary models \cite{Anber:2009ua,Barnaby:2012xt,Anber:2012du,Domcke:2016bkh}, the seeding of cosmological magnetic fields \cite{Garretson:1992vt, Ratra:1991bn,Anber:2006xt,Choi:2018dqr,Fujita:2015iga,Adshead:2016iae}, reducing the relic abundance of the QCD axion \cite{Kitajima:2017peg,Agrawal:2017eqm}, populating vector dark matter \cite{Dror:2018pdh,Co:2018lka,Bastero-Gil:2018uel,Agrawal:2018vin}, friction for the relaxion mechanism \cite{Hook:2016mqo,Fonseca:2018xzp}, and GW from the string axiverse~\cite{Soda:2017dsu,Kitajima:2018zco}.
Here, we assume that the axion starts rolling sometime after the end of inflation when the universe is radiation dominated, as is the case in relaxion~\cite{Hook:2016mqo,Fonseca:2018xzp} or axion-curvaton models. The resulting GW signal will therefore be peaked, with the peak frequency set by the axion mass and and the amplitude by the Hubble rate at the time when the backreaction from the tachyonic instability becomes large. 

As shown in Figure~\ref{fig:money_plot}, the tail of the GW spectrum from the QCD axion may be detectable by pulsar timing arrays while generic ALP models are in reach of LISA and other future GW detectors. At the same time, the relic abundance of the axion can be a fraction or all of the observed DM in the universe. 

This paper is organized as follows: After describing the model and initial conditions in Section \ref{s:model}, we review and discuss the particle production mechanism in Section \ref{s:darkphoton}. Section \ref{s:gws} contains an outline of the GW spectrum computation and gives an analytic estimate for the peak frequency and amplitude. Plots of the GW spectra from our numerical simulation can be found in Section \ref{s:results}. 
%
\section{Model}
\label{s:model}
Our model consists of an axion $\phi$ and a dark photon $X_{\mu}$ of an unbroken $U(1)_X$ gauge symmetry under which the Standard Model fields are uncharged~\footnote{We assume there are no light degrees of freedom which carry $U(1)_X$ charge, otherwise exponential production of the dark vector may be impeded by the resulting Debye mass.}\footnote{We define $\widetilde{X}^{\mu\nu}=\epsilon^{\mu \nu \alpha \beta} X_{\alpha \beta}/2$ with $\epsilon^{0123}=1/{\sqrt{-g}}$.}
\begin{equation}
\begin{split}
\mathcal{S} = \int d^{4}x \,\sqrt{-g} \, \bigg[ \frac{1}{2}\partial_{\mu} \phi \,\partial^{\mu} \phi - V(\phi) & \\
-\frac{1}{4} X_{\mu\nu} X^{\mu\nu} - \frac{\alpha}{4f}\phi X_{\mu\nu}\widetilde{X}^{\mu\nu} \bigg] & \, ,
\end{split}
\label{lag}
\end{equation}
where $f$ is the scale of the global symmetry breaking that gives rise to the Nambu-Goldstone field $\phi$. 
We assume a cosine-like potential for the axion with mass $m$
\begin{equation}
V(\phi) = m^{2}f^{2}\left[ 1- \cos\left( \frac{\phi}{f}\right)\right] \, ,
\label{phi_pot}
\end{equation}
however our results do not depend crucially on the precise form of the potential, allowing the mechanism to be extended to other types of rolling pseudoscalars.
Adopting a metric convention of $ds^{2} = a(\tau)^{2} (d\tau^{2} - \delta_{ij}dx^{i}dx^{j})$ and assuming that $\phi$ is spatially homogeneous, the equation of motion for the axion field is
\begin{equation}
\phi'' + 2aH\phi' + a^{2}\frac{\partial V}{\partial \phi} =  \frac{\alpha}{f} a^{2} \vec{E}\cdot\vec{B} \,,
\label{phi_eom}
\end{equation}
where primes indicate derivatives with respect to the conformal time $\tau$, $\vec{E}$ and $\vec{B}$ are the dark electric and magnetic fields, and the Hubble parameter is defined as $H = a'/a^{2}$. As the axion rolls towards its minimum, the coupling $\phi X_{\mu\nu}\widetilde{X}^{\mu\nu}$  can lead to exponential production of $X_\mu$ quanta. 

Different from most of the existing literature, here we assume that the dynamics takes place after inflation, in a radiation dominated epoch. More precisely, we assume the following \textbf{initial conditions}, which can naturally arise at the end of inflation:
\begin{itemize}
	\item $\phi$ is displaced from its minimum by $\Delta\phi = \theta f$ with the initial misalignment angle $\theta = {\cal O}(1)$.
	\item The energy density in $\phi$ is smaller than that of the radiation bath, such that its backreaction on the geometry can be ignored.
	\item The gauge field $X_{\mu}$ is not thermalized and thus has zero initial abundance.
	\item The initial velocity $\phi'$ is negligible.
\end{itemize}
The last two assumptions may be relaxed without spoiling our mechanism. 

While $H \gg m$, the axion is pinned by Hubble friction and no gauge bosons are produced. At the temperature $T_{\rm osc}$ defined by $H_{\rm osc} \approx m$, the axion becomes free to roll toward the minimum of its potential. In a radiation dominated universe, the Hubble expansion rate is approximately $H \approx T^{2}/M_{P}$, so the temperature at which the axion begins to oscillate is $T_{\rm osc} \approx \sqrt{m M_P}$, where $M_P$ is the reduced Planck scale.

The rolling axion induces a tachyonic instability in the gauge boson equation of motion. This allows dark photon modes in a specific frequency band to grow exponentially, amplifying quantum fluctuations of $X_{\mu}$ into classical modes and transferring a large fraction of the axion energy density into dark radiation. It is this process of exponential particle production amplifying quantum fluctuations within a characteristic frequency band (or set of length scales) that we later identify as the source of gravitational radiation. 
\section{Dark Photon Production}
\label{s:darkphoton}
We now review the particle production mechanism more concretely, following Refs.~\cite{Anber:2009ua,Barnaby:2012xt,Anber:2012du,Domcke:2016bkh}. Working in the Coulomb gauge defined by $\vec{\nabla}\cdot \vec{X} = 0$, the equation of motion for $\vec{X}$ is
\begin{equation}
\left( \frac{\partial^{2}}{\partial \tau^{2}} - \nabla^{2} -  \alpha\frac{\phi'}{f}\vec{\nabla}\times\right)\vec{X} = 0 \,.
\end{equation}
To study the production of dark photons, we quantize the dark gauge field as
\begin{equation}
\begin{split}
&\hat{X}^{i}({\bf x}, \tau) =  \int \frac{d^{3}k}{(2\pi)^3}\hat{X}^{i} ({\bf k}, \tau) e^{i{\bf k} \cdot {\bf x}} \\
 &= \!\sum_{\lambda=\pm}\! \int\! \frac{d^{3}k}{(2\pi)^3} v_{\lambda} (k, \tau)  \, \veps^{i}_{\lambda}({\bf k})\, \hat{a}_{\lambda}({\bf k})\, e^{i{\bf k} \cdot {\bf x}}\!+\text{h.c.} \,,
\end{split}
\end{equation}
where $[\hat{a}_{\lambda}({\bf k}),\hat{a}^{\dagger}_{\lambda'}({\bf k'})] = (2\pi)^{3} \delta_{\lambda\lambda'} \delta({\bf k-k}')$ and the circular polarization vectors satisfy ${\bf k} \cdot \bveps^{\pm} = 0$, ${\bf k} \times \bveps^{\pm} = \mp i k \bveps^{\pm}$, $\bveps^{\pm} \cdot \bveps^{\pm} = 0$, and $\bveps^{\pm} \cdot \bveps^{\mp} = 1$. It follows that the dark photon mode functions $v_{\lambda} (k, \tau)$ satisfy
\begin{equation}
v''_{\pm} (k, \tau)+ \omega^{2}_{\pm}(k, \tau) \, v_{\pm}(k, \tau) = 0 \,,
\label{eq:mode_eq}
\end{equation}
with a time-dependent frequency
\begin{equation}
\omega^{2}_{\pm}(k, \tau)= k^{2}  \mp k\frac{\alpha}{f} \phi'\,.
\end{equation}
As $\phi$ begins to roll, one of the helicities 
will have negative values of $\omega^2$ for modes in the range $0 < k < \alpha |\phi'|/f$. This corresponds to a tachyonic instability which causes the corresponding dark photon helicity to grow like $v_{\pm} \sim e^{|\omega_{\pm}| \tau}$. 
The fastest growing mode is $\tilde{k} = \alpha |\phi'|/(2f)$, where $\tilde{\omega}^{2}(\tilde{k}) = -\tilde{k}^{2}$ is the most negative tachyonic frequency. \emph{Since the mode with momentum $\tilde{k}$ grows the fastest and has the most energy, it will set the peak of the gauge and gravitational wave power spectra.} 

We numerically solve the equations of motion and calculate the GW spectrum in Section \ref{s:results}. However, let us first offer some analytic understanding of the dynamics which determine the shape and amplitude of the GW spectrum. 

The dependence of $\tilde{k}$ on $\phi'$ means that the scale where most of the energy is being deposited becomes larger as the system evolves. To understand the evolution of $\tilde{k}$, we approximate the solution of Eq.~\ref{phi_eom} in physical time $t$ as
\begin{equation}
\phi(t) = \phi_{\rm osc}\left(\frac{a_{\rm osc}}{a}\right)^{\frac{3}{2}} \cos(mt) \,,
\label{eq:phi_t}
\end{equation}
which holds while the friction from production of dark photons is small. With this, we can approximate the envelope of $\phi'$ as
\begin{equation}
|\phi'| = \bigg| \frac{d\phi}{dt}\frac{dt}{d\tau}\bigg|  \approx \theta f \left(\frac{a_{\rm osc}}{a}\right)^{\frac{3}{2}} am\,,
\end{equation}
which then allows us to write the scale $\tilde{k}$ in terms of Lagrangian parameters and the scale factor
\begin{equation}
\tilde{k} \approx \frac{\alpha\theta}{2} \left(\frac{a_{\rm osc}}{a}\right)^{\frac{3}{2}} am \, .
\end{equation}
\subsection{Tachyonic Production Band}
It is important to point out that since the helicity which experiences tachyonic instability flips between $``+"$ and $``-"$ when $\phi'$ changes sign every half period, no modes experience significant growth unless their growth timescale $|\omega|^{-1}$ is less than the conformal oscillation time $(am)^{-1}$. We define the tachyonic production band $k_{-} < k < k_{+}$ as the range of modes which are both tachyonic and have growth times less than $(am)^{-1}$, where $k_{-}$ and $k_{+}$ are given by solving $\omega^{2} = -(am)^{2}$. The result is
\begin{equation}
k_{\pm}\approx \tilde{k} \left[ 1\pm \sqrt{1-\left(\frac{2}{\alpha\theta}\right)^{2}\left(\frac{a}{a_{\rm osc}}\right)^{3}}\,\right]\,,
\label{eq:TPB}
\end{equation}
from which we see that $\alpha\theta > 2$ is required to have the tachyonic production band open initially. As the universe expands, the band contracts while simultaneously shifting toward lower $k$ like $\tilde{k} \propto a^{-1/2}$. The band closes when $(a/a_{\rm osc}) = (\alpha\theta/2)^{2/3}$ where  $\tilde{k} = a_{\rm osc}\, m\, (\alpha\theta/2)^{2/3}$. As an example, to keep the band open until the scale factor has grown by a factor of $\sim5$, one requires $\alpha\theta \gtrsim 20$.\ca{\footnote{As pointed out in Ref.~\cite{Agrawal:2017eqm}, a possible way to achieve $\alpha >1$ is using the alignment mechanism~\cite{Kim:2004rp}. A detailed discussion can also be seen in Ref.~\cite{Agrawal:2018mkd,Agrawal:2018vin}.}} We will later use the value of the scale factor when the tachyonic band closes to estimate the scale $\tilde{k}$ at the time of gravitational wave production.
\subsection{Parity Violation and Dark Photon Chirality}
\label{sec:dppv}
As was pointed out in Ref.~\cite{Sorbo:2011rz}, the gauge field helicities are not produced in equal amounts because the operator $\alpha \phi X \widetilde{X}/f$ violates parity when ${\langle \phi\rangle \neq 0}$. This parity violating effect manifests itself as friction due to production of dark photons of a single helicity. From misalignment arguments we expect $\langle\phi\rangle =\theta f$ initially, so the amount of parity violation is controlled by $\alpha\theta$.  It also follows that the initial sign of $\phi'$ and thus the first helicity to become tachyonic (without loss of generality we take this to be ``+") are randomly selected. 

As $\phi$ rolls, the amplitude of $\phi'$ decays due to friction from the expansion of the universe and particle production. As a result, when $\phi'$ changes sign and the opposite helicity becomes tachyonic, it receives dramatically less enhancement since the growth of the mode functions depends exponentially on $\phi'$. 

\begin{figure}
\centering
\includegraphics[width=0.95 \columnwidth]{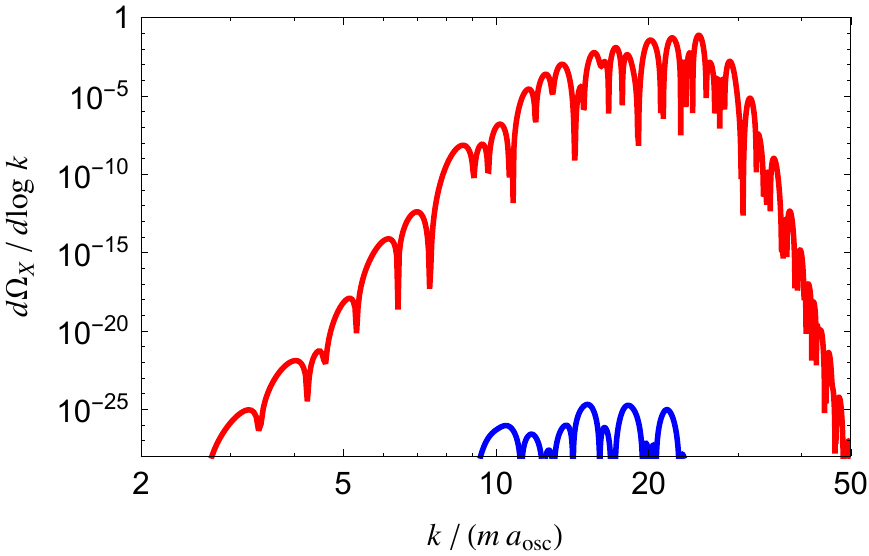}
\caption{Dark photon spectral energy density when the tachyonic band has closed for model parameters $m = 10$ meV, $f=10^{17}$ GeV, $\alpha =55$, and $\theta = 1.2$. The red line corresponds to the spectrum for $``+"$ helicity dark photons while the blue line  gives the $``-"$ helicity spectrum.}
\label{fig:dp_pols}
\end{figure}

In Figure~\ref{fig:dp_pols} we show the dark photon spectral energy density after the tachyonic band has closed, where the exponential suppression of one helicity with respect to the other can be clearly seen. The broad peak of the spectrum is generated as $\tilde{k}$ evolves toward larger scales. Thus, the value of $\tilde{k}$ when the axion begins to oscillate gives the right edge of the peak and its value when the tachyonic band closes roughly gives the left edge. As we discuss later and in Appendix \ref{appB}, the final spectrum also features additional enhancement from parametric resonance after the tachyonic band has closed.
%
\section{Gravitational Waves}
\label{s:gws}
To study gravitational waves, we consider the perturbed metric
\begin{equation}
ds^{2} = a(\tau)^{2} \left[ d\tau^{2} - (\delta_{ij}+h_{ij}) dx^{i}dx^{j}\right] \, .
\end{equation}
The linearized Einstein equations for $\bar{h}_{ij} \equiv a h_{ij}$ in Fourier space are 
\begin{equation}
\bar{h}''_{ij}({\bf k},\tau) +\left(k^{2} -\frac{a''}{a}\right) \bar{h}_{ij}({\bf k},\tau)= \frac{2a}{M_{P}^{2}}\Pi_{ij}({\bf k},\tau)\, , 
\end{equation}
where $\Pi_{ij}({\bf k},\tau) = \Lambda_{ij}^{kl}T_{kl}({\bf k},\tau)$ is the anisotropic part of the energy-momentum tensor $T_{ij}$. The object $\Lambda_{ij}^{kl} = \Lambda_{i}^{k} \Lambda_{j}^{l} - \frac{1}{2}\Lambda_{ij}\Lambda^{kl}$ is the transverse traceless projector with $\Lambda_{ij} = \delta_{ij} -k_{i}k_{j}/k^{2}$. Since gravitational waves are sourced by the highly occupied modes of the dark photon, the quantity $\Pi_{ij}$ is an operator
\begin{align}
\hat{\Pi}_{ij}({\bf k},\tau) = & -\frac{\Lambda_{ij}^{kl} }{a^{2}} \int \frac{d^{3}q}{(2\pi)^3}  \Big[ \hat{E}_{k}({\bf q},\tau)\hat{E}_{l}({\bf k-q},\tau) \notag \\
 &+ \hat{B}_{k}({\bf q},\tau)\hat{B}_{l}({\bf k-q},\tau) \Big] \, , \label{eq:piij}
\end{align}
where
\begin{align}
\hat{E}_{i}({\bf q},\tau) &= \hat{X}'_{i}({\bf q},\tau) = v'_{\lambda} (q, \tau)  \, \veps_{i}^{\lambda}({\bf q})\, \hat{a}_{\lambda}({\bf q})\,, \notag\\
\hat{B}_{i}({\bf q},\tau) &= -i \epsilon_{ijk}q_{j}\hat{X}_{k}({\bf q},\tau) \notag\\
&= \lambda \, q\, v_{\lambda} (q, \tau)\,  \veps_{i}^{\lambda}({\bf q})\, \hat{a}_{\lambda}({\bf q})\,.\notag
\end{align}
Neglecting the $a''$ term in the equation of motion for $h_{ij}$ (which vanishes in a radiation dominated universe where $a\propto \tau$), the solution for $h_{ij}$ is
 \begin{equation}
\hat{h}_{ij}({\bf k},\tau)  = \frac{2}{aM_{P}^{2}} \int_{\tau_{\rm osc}}^{\tau} d\tau' a(\tau') \mathcal{G}(k,\tau,\tau') \hat{\Pi}_{ij}({\bf k},\tau')  \, ,
\label{eq:hij_hat}
\end{equation}
where $\mathcal{G}= \sin\left[ k(\tau-\tau')\right]/k $ is the causal Green's function for the d'Alembert operator.
The gravitational wave spectral energy density is given by
\begin{equation}
\frac{d\rho_{\rm GW}}{d \log k} = \frac{M_{P}^{2}k^{3}}{8\pi^{2}a^{2}} \mathcal{P}_{h'}({\bf k},\tau) \, ,
\end{equation}
with the power spectrum $\mathcal{P}_{h'}({\bf k},\tau)$ defined by $ \langle 0| \hat{h}'_{ij}({\bf k},\tau)\hat{h}^{'*}_{ij}({\bf k}',\tau) |0 \rangle = (2\pi)^{3}\mathcal{P}_{h'}({\bf k},\tau) \delta({\bf k}-{\bf k}')$. Inserting the solution for $\hat{h}_{ij}$, we find
\begin{equation}
\begin{split}
\frac{d\rho_{\rm GW}}{d \log k} =&\frac{k^{3}}{4\pi^{2}a^{4}M_{P}^{2}}\int_{\tau_{\rm osc}}^{\tau} d\tau' d\tau'' a(\tau')a(\tau'') \\
&\times \cos\left[k(\tau'-\tau'') \right]\Pi^{2}({\bf k},\tau',\tau'') \, ,
\end{split}
\label{drhobydlogk}
\end{equation}
where we have averaged over one period $\Delta\tau = 2\pi/k$ which gives a factor of 1/2. The function $\Pi^{2}({\bf k},\tau',\tau'') $ is the Unequal Time Correlator and is defined as $ \langle 0| \hat{\Pi}_{ij}({\bf k},\tau)\hat{\Pi}^{*}_{ij}({\bf k}',\tau') |0 \rangle = (2\pi)^{3}\Pi^{2}({\bf k},\tau,\tau') \delta({\bf k}-{\bf k}')$. A full calculation of this object and the final formula for the gravitational wave spectrum in terms of the dark photon mode functions can be found in Appendix~\ref{app:first_app}. Useful for comparison to experiment is the fractional gravitational wave spectral energy density defined by
\begin{equation}
\Omega_{\rm GW}(k) \equiv  \frac{1}{\rho_{\rm tot}}\frac{d\rho_{\rm GW}}{d \log k} \, ,
\end{equation}
the value of which today is usually plotted as $h^{2}\Omega_{\rm GW}^{0}$ with $h = H_{0}/100$ and $H_{0} = 67.8 \,\, {\rm km\,s^{-1}\,Mpc^{-1}}$.
\subsection{Estimating the Gravitational Wave Spectrum}
The energy density stored in the axion field at $T_{\rm osc}$, 
\begin{align}
	\Omega_{\phi}^{\rm osc} = \frac{\rho_{\phi}^{\rm osc}}{\rho_{\rm tot}^{\rm osc}} \approx \frac{m^{2}\theta^{2}f^{2}/2}{3M_{P}^{2}H_{\rm osc}^{2}} \approx \left(\frac{\theta f}{M_P}\right)^{2}\,,
\label{eq:omega_osc}
\end{align}
sets an upper bound for the amount of energy that can be transformed into gravitational radiation.
Notice that large decay constants are required to have an observable GW signal in planned future experiments. Normally, for $f\gtrsim 10^{17}$~GeV and $\theta \sim \mathcal{O}(1)$, the axion mass should be less than about $10^{-23}$ eV so that the relic abundance does not overshoot the observed dark matter density. However, the efficient transfer of axion energy density into dark radiation provided by the tachyonic instability and subsequent phase of parametric resonance reopens this parameter space.

The energy taken from $\phi$ during the tachyonic phase of particle production is transferred to dark photon modes in the range $k_{-}<k<k_{+}$ with the peak energy deposition occuring at the scale $\tilde{k}$. Therefore, we expect $\tilde{k}$ at the time of gravitational wave emission (which we denote as $\tilde{k}_{*}$) to set the location of the peak of the gravitational wave spectrum
\begin{equation}
k_{\rm peak}\approx 2 \tilde{k}_{*} \approx  \alpha\theta \left(\frac{a_{\rm osc}}{a_{*}}\right)^{\frac{3}{2}} ma_{*} \,,
\label{eq:kpeak}
\end{equation}
where the factor of 2 approximates the addition of dark photon momenta. We use the value of the scale factor when the tachyonic band closes $a/a_{\rm osc} = (\alpha\theta/2)^{2/3}$ to estimate the scale factor $a_{*}$ at the time of gravitational wave emission. With this, we estimate the location of the peak of the GW spectrum as
\begin{equation}
k_{\rm peak}\approx  (\alpha\theta)^{2/3} \,m\, a_{\rm osc}\, .
\label{eq:kpeak}
\end{equation}
Following Refs. \cite{Giblin:2014gra,Buchmuller:2013lra}, we postulate a simple scaling relation for the peak amplitude of the gravitational wave spectrum at the time of emission
\begin{equation}
\Omega_{\rm GW}(k_{\rm peak}) = \Omega_{s}^{2} \, \left( \frac{a_{*}H_{*}}{k_{\rm peak}}\right)^{2}  \,,
\label{eq:scal_rel}
\end{equation}
where $(a_{*}H_{*})^{-1}$ is the comoving horizon at the time of emission and we have defined the energy density fraction of the gravitational wave source as $\Omega_{s} = c_{\rm eff} \, \Omega_{\phi}^{*}$. Here, $c_{\rm eff}$ is a model dependent factor characterizing the efficiency of converting energy in the source to gravitational waves. 
Causality gives an upper bound on the gravitational wave amplitude, since $k_{\rm peak} < a_{*} H_{*}\leq a_{*}m$ corresponds to all modes having longer growth timescales than $(ma_{*})^{-1}$, so there is no tachyonic particle production. In the radiation dominated era, we have $\Omega_{\phi}^{*}/\Omega_{\phi}^{\rm osc} = a_{*}/a_{\rm osc}$ and $H_{*}/H_{\rm osc} = (a_{\rm osc}/a_{*})^{2} \approx H_{*}/m$. Using Eq.~\ref{eq:omega_osc} we write the scaling relation Eq.~\ref{eq:scal_rel} in terms of the model parameters at the time when the axion begins to oscillate as 
\begin{equation}
\Omega_{\rm GW}(k_{\rm peak})  \approx c_{\rm eff}^{2} \left( \frac{f}{M_P} \right)^4 \, \left( \frac{\theta^{2}}{\alpha}\right)^{\frac{4}{3}}  \,.
\label{eq:SRest}
\end{equation}
We note that our estimate here gives the contribution to the peak amplitude coming only from the tachyonic phase of particle production. The peak amplitude is further enhanced as the physical momemtum $k_{\rm peak}/a$ redshifts and enters the narrow parametric resonance band (see Appendix \ref{appB}). Thus, there is reason to expect this scaling relation to underestimate the actual peak amplitude.

\subsection{Present Time Gravitational Wave Spectrum}
To obtain the amplitude and frequency of the gravitational wave spectrum today, we need to account for redshifting. The emitted amplitude $\Omega_{\rm GW}^*$ is redshifted by a factor
\begin{equation}
\Omega_{\rm GW}^{0}  =\Omega_{\rm GW}^{*}  \left(\frac{g_{s,{\rm eq}}}{g_{s,*}}\right)^{\frac{4}{3}} \left(\frac{T_{\rm 0}}{T_{*}}\right)^{4} \left(\frac{H_{*}}{H_{0}}\right)^{2} \, ,
\end{equation}
with $g_{s,{\rm eq}} = 2+2N_{\rm eff}(7/8)(4/11) = 3.938$ and $T_0=2.73$~K. Assuming radiation domination at the time of emission, the amplitude today can also be written as
\begin{equation}
\begin{split}
\Omega_{\rm GW}^{0}  &=\Omega_{\rm GW}^{*}  \left(\frac{g_{s,{\rm eq}}}{g_{s,*}}\right)^{\frac{4}{3}} \left( \frac{g_{\rho,*}}{g_{\rho,0}^{\gamma}}\right) \Omega_{\gamma}^{0}\, \\
&\approx 1.67\times10^{-4} \, g_{\rho,*}^{-1/3} \, \Omega_{\rm GW}^{*} ,
\end{split}
\label{eq:amp_RS_simple}
\end{equation}
where $g_{\rho}$ is the number of effective degrees of freedom in the thermal bath associated with the energy density and we have $g_{\rho,0}^{\gamma} = 2$ and $\Omega_{\gamma}^{0} = 5.38\times 10^{-5}$ \cite{Tanabashi:2018oca}. Additionally, for the last step we have made the approximation $g_{s,*} = g_{\rho,*}$, which is very good for emission temperatures $T_{*} > m_e$. 

The physical peak frequency redshifts as $f_{\rm peak} = k_{\rm peak} / a$, so its value today is given by
\begin{equation}
 f_{0}^{\rm peak} = \frac{ k_{\rm peak} }{a_{0}}= \left(\frac{g_{s,{\rm eq}}}{g_{s,{\rm osc}}}\right)^{\frac{1}{3}} \left(\frac{T_{\rm 0}}{T_{\rm osc}}\right) \frac{ k_{\rm peak} }{a_{\rm osc}}\, .
\end{equation}
Inserting $k_{\rm peak}$ from Eq.~\ref{eq:kpeak}, we see that the peak frequency is related to the axion mass via
\begin{align}
	f_{0}^{\rm peak} &\approx (\alpha\theta)^{\frac{2}{3}} \,T_{0} \,  \left(\frac{g_{s,{\rm eq}}}{g_{s,\rm osc}}\right)^{\frac{1}{3}}  \left(\frac{m}{M_{P}}\right)^{\frac{1}{2}} \notag \\
	& \approx 6\times10^{-4}\,\, {\rm Hz} \, \left( \frac{\alpha\theta}{66} \right)^{\frac{2}{3}} \left( \frac{m}{10 \,{\rm meV}} \right)^{\frac{1}{2}}.
	\label{eq:peak_freq}
\end{align}

\section{Results}
\label{s:results}
We numerically solve the coupled axion and dark photon equations of motion, treating the backreaction from particle production by assuming the axion responds to the expectation value of $\vec{E}\cdot \vec{B}$
\begin{equation}
\begin{split}
 & \vec{E}\cdot\vec{B} \rightarrow \langle 0|\vec{E}\cdot\vec{B}|0\rangle \\
  &=\frac{1}{2\pi^2 a^4}\sum_{\lambda=\pm} \lambda \int dk\,  k^3\, {\rm Re}[v_\lambda(k,\tau)v^{\prime*}_\lambda(k,\tau)] \, .
 \label{RHS_phi_eom}
 \end{split}
\end{equation}
Since we assume that the axion field is homogeneous, the equation of motion for the gauge modes only depends on $|\vec{k}|=k$. The mode functions $v_\lambda(k,\tau)$ are included in Eq.~\ref{RHS_phi_eom} by discretizing the momenta $k$ and approximating the integral as a sum over simulated modes. We discretize using 5000 equally spaced modes with momenta ranging from $0$ to $\theta\alpha ma_{\rm osc}$, where the mode functions were initially taken to be in the Bunch-Davies vacuum $v_{\lambda}(k,\tau \ll \tau_{\rm osc}) = e^{-ik\tau}/\sqrt{2k}$. We start the simulation at the temperature defined by $H = m$ and integrate until energy transfer has ended. All of our benchmark points have $\Omega_{\phi},\Omega_{X} < 1$, so the total energy density of the universe is dominated by radiation and we neglect the energy density in the axion and dark photon when calculating the background evolution. All changes in the number of relativistic degrees of freedom are fully taken into account following Ref.~\cite{Husdal:2016haj}. We assume no temperature dependence of the axion mass, which is a good approximation when $f\gtrsim10^{17}$ GeV in case of the QCD axion. 

\subsection{Benchmark Gravitational Wave Spectra}
\begin{figure*}[t]
\centering
\includegraphics[width=1.0\textwidth]{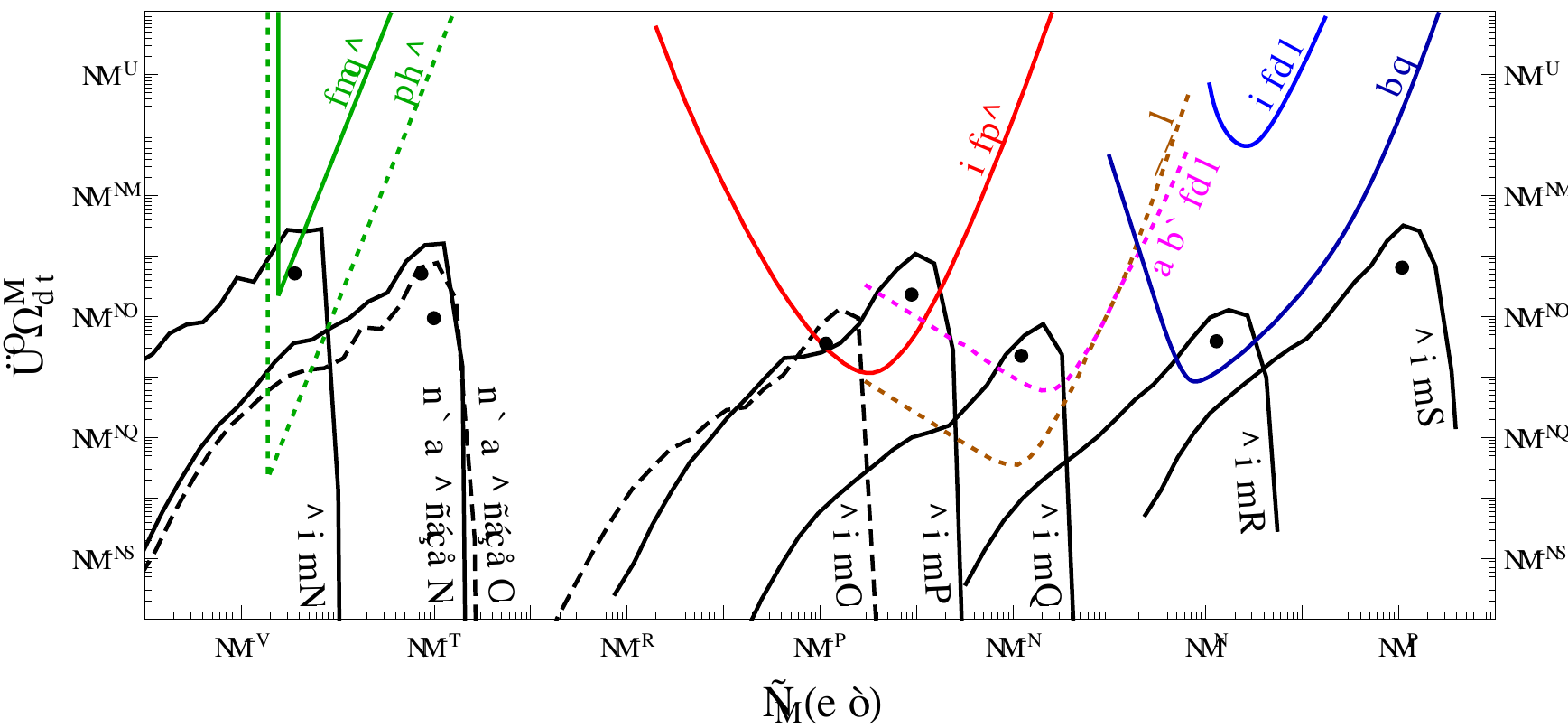}
\caption{Black lines give benchmark gravitational wave spectra at the present time for various values of the model parameters (shown in Table~\ref{tab:GW_spec_params}). The black dots show the prediction of the peak location using the scaling relation in Eq.~\ref{eq:SRest}. Colored curves are projected power law sensitivities for various gravitational wave detectors. Green (dotted):  IPTA (SKA), Red: LISA 4-yr, Blue: LIGO 2022, Brown: DECIGO, Magenta: BBO, Dark Blue: Einstein Telescope. }
\label{fig:money_plot}
\end{figure*}

To compute the gravitational wave spectra, we express Eq.~\ref{drhobydlogk} in terms of the simulated mode functions (see Appendix~\ref{app:first_app} for details). Discretizing the resulting double integral over momenta results in the computation time growing as $\mathcal{O}(N^{2})$, so these integrals were computed using a subset of $N = 100$ of the total 5000 simulated modes. We checked that increasing the number of modes produced no significant changes in our results. We computed the gravitational wave spectra for several different sets of model parameters shown in Table~\ref{tab:GW_spec_params} and the results are shown in Figure~\ref{fig:money_plot}.

Both peak and amplitude of the numerically obtained GW spectra agree reasonably well with the analytic estimates. The signals are detectable in LISA and current PTA experiments if the peak falls into the most sensitive regions of the experiments. Future experiments with sensitivities significantly below $h^2\Omega_{\rm GW}^{0}  \sim 10^{-13}$ could even detect the tails of the GW signals and thus probe larger bands of axion masses. In particular, SKA could observe a GW signal from the QCD axion. Since the axion dark matter abundance is very sensitive to small variations of the initial conditions, we only demand that our benchmark points do not grossly overproduce dark matter. Therefore all the points listed in Table~\ref{tab:GW_spec_params} should be considered consistent with cosmology. 

It is intriguing that some of the parameter space for axion dark matter might first be probed by GW detectors. The low mass region $10^{-19}~\text{eV} \lesssim m \lesssim 10^{-13}~\text{eV}$ will be probed indirectly by the black hole superradiance with data from LISA~\cite{Cardoso:2018tly}, showing some unexpected complementarity of GW measurements by LISA and PTAs. 

\begin{table}[h!]
\renewcommand{\arraystretch}{1.25}
\begin{center}
\resizebox{\columnwidth}{!}{%
\begin{tabular}{|c|c|c|c|c|c|c|}
\hline \hline
GW Spectrum   &  $m$ (eV) & $f$ (GeV) & $\theta$ & $\alpha$ & $\rho_{\phi}^{0}/\rho_{\rm DM}^{0}$ & $\Delta N_{\rm eff}$ \\
\hline 
ALP 1   &  $5.6\times10^{-14}$ & $2.0\times 10^{17}$ & 1.0 & 75 & $0.011$ & $0.24$       \\ 
QCD Axion 1   &  $3.0\times10^{-11}$  & $2.0\times10^{17}$ & 1.0 & 73 & $1.1$ & $0.18$  \\ 
QCD Axion 2   &  $6.1\times10^{-11}$  & $1.0\times10^{17}$ & 1.3 & 55 & $1.9$ & $0.075$  \\ 
ALP 2   &  $1.0\times10^{-2}$ & $1.0\times 10^{17}$ & 1.2 & 55 & $1.7$ & $0.030$       \\ 
ALP 3   &  $5.0\times10^{-1}$ & $2.0\times 10^{17}$ & 1.0 & 75 & $0.85$ & $0.069$       \\ 
ALP 4  &  $1.0\times10^{2}$ & $1.0\times 10^{17}$ & 1.1 & 65 & $0.020$ & $0.018$       \\ 
ALP 5  &  $1.0\times10^{6}$ & $1.0\times 10^{17}$ & 1.3 & 60 & $0.29$ & $0.020$       \\ 
ALP 6  &  $1.0\times10^{10}$ & $2.0\times 10^{17}$ & 1.2 & 50 & $*$ & $*$       \\ 
  \hline \hline
\end{tabular}
}
\end{center}
\caption{Parameter values for the gravitational wave spectra shown in Figure~\ref{fig:money_plot}. The present time ratio of the axion and DM energy densities is given by $\rho_{\phi}^{0}/\rho_{\rm DM}^{0}$, except for the last benchmark point where the axion is not cosmologically stable.  } 
\label{tab:GW_spec_params}
\end{table}
\subsection{Chirality of the Gravitational Wave Spectrum}
\begin{figure}
\centering
\includegraphics[width=0.95 \columnwidth]{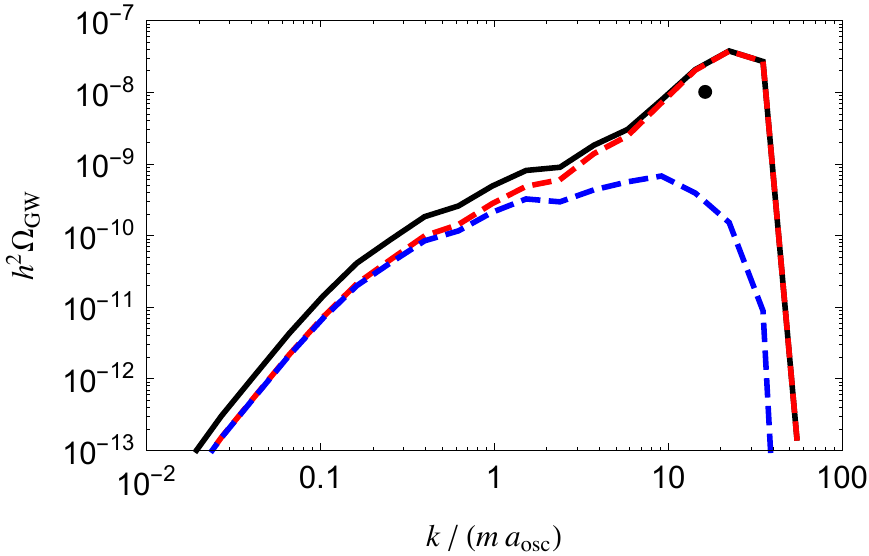}
\caption{Emission time gravitational wave spectrum for the ALP 2 model parameters. The solid black line gives the total spectrum while the dashed lines show the contributions from the $``+"$ (red) and $``-"$ (blue) helicities of the spectrum.}
\label{fig:GW_pols}
\end{figure}

As we discussed in Section~\ref{sec:dppv}, the dark photon population is completely dominated by a single helicity and has a relatively narrow range of momenta corresponding to the modes that experienced significant tachyonic growth.
Since gravitational waves are sourced by exponentially amplified dark photon quantum fluctuations, they inherit the parity violation in the dark photon population.  The peak of the gravitational wave spectrum comes from the addition of two approximately parallel ``+" polarized dark photons of similar momenta $k$, such that a $``+"$ circularly polarized gravitational wave is produced with momentum $\approx 2k$. In contrast, the low-$k$ tail of the gravitational wave spectrum comes from two approximately anti-parallel ``+" polarized dark photons of similar momenta $k$. This results in an approximate cancellation of the polarizations and momenta, leading to the production of unpolarized, low momentum gravitational waves. 
These features can be seen in Figure~\ref{fig:GW_pols}, where the peak of the gravitational wave spectrum is dominated by $``+"$ polarized gravitational waves while the tail has equal components of both helicities such that the net spectrum is unpolarized.
\subsection{Relic Abundance and $N_{\rm eff}$}
\begin{figure}
\centering 
\includegraphics[width=0.95 \columnwidth]{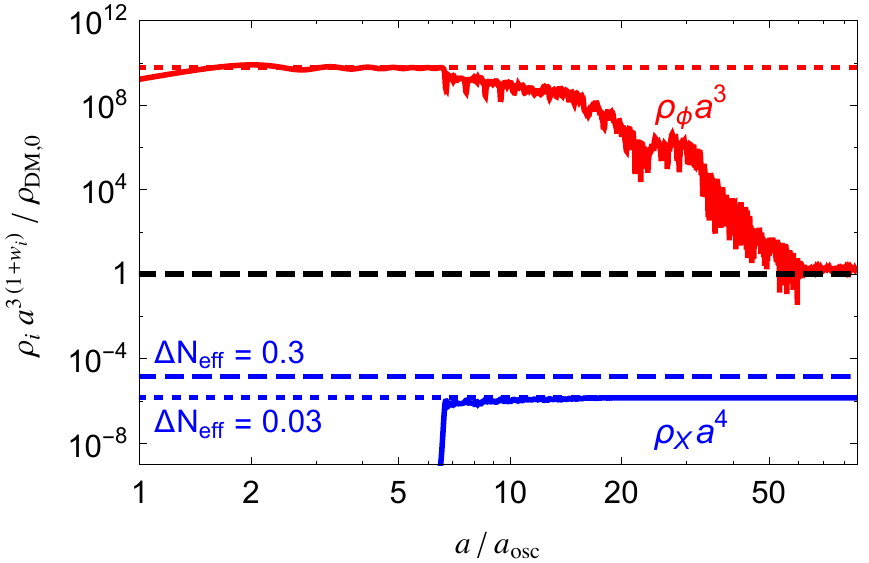}
\caption{Comoving energy densities normalized to the present time dark matter energy density, for the ALP 2 benchmark model. Here, the scale factor is normalized to unity at the present time.}
\label{fig:CME_and_GPS}
\end{figure}

The dark photon modes that become highly occupied in the tachyonic instability phase also allow for further efficient transfer of energy from the axion to the gauge fields via parametric resonance. A detailed discussion of the latter can be found in Appendix~\ref{appB}. In combination, the two mechanisms can suppress the axion DM abundance by up to 14 orders of magnitude, significantly widening the range of axion masses and decay constants that are consistent with cosmology~\cite{Agrawal:2017eqm}. 
As an example, the red curve in Figure~\ref{fig:CME_and_GPS} shows the suppression of the axion abundance compared to the case with no particle production for the benchmark model ALP 2.

The energy density of the dark photons dilutes as radiation and changes the number of effective relativistic degrees of freedom ($N_{\rm eff}$). At the recombination time, the dark radiation contribution to $N_{\rm eff}$ is given by
\begin{equation}
\Delta N_{\rm eff} = \frac{8}{7}\left(\frac{11}{4}\right)^{\frac{4}{3}} \frac{\rho_{X}}{\rho_{\gamma}}\bigg|_{T=T_{\rm rec}} \,,
\end{equation}
where the energy density of photons is $\rho_{\gamma}$ and of dark photons is $\rho_{X}$. The Planck 2018 TT,TE,EE,lowE+lensing+BAO dataset constrains $\Delta N_{\rm eff} < 0.3$ at 95\% confidence level~\cite{Aghanim:2018eyx}. The changes in $N_{\rm eff}$ generated by our benchmark points, shown in the right most column of Table~\ref{tab:GW_spec_params}, are largely consistent with this bound, with some points having a mild $\sim 1\sigma$ tension. The next generation of ground-based telescope (CMB Stage-4) experiments conservatively expect to achieve a sensitivity of $\Delta N_{\rm eff} < 0.03$~\cite{Abazajian:2016yjj}, which is sufficient to probe much of the parameter space which gives a large gravitational wave signal.

If the axion is heavy, decays to dark photons will deplete the relic abundance. The axion decay width to dark photons is given by
\begin{equation}
\Gamma_{\phi \rightarrow XX} = \frac{\alpha^{2}m^{3}}{64\pi f^{2}} \,,
\end{equation}
so for $f\sim 10^{17}$ GeV and $\alpha \sim 50-100$ the axion decays before Big Bang Nucleosynthesis for ${m \gtrsim 10^{3}}$~GeV and before recombination for $m \gtrsim 10$ GeV. The GW peak frequency scales with the axion mass as in Eq.~\ref{eq:peak_freq}, which for $m \gtrsim 10^{3}$ (10) GeV translates into $f_{0} \gtrsim 10^{4}$ ($10^{3}$) Hz.

We note that in this work we ignore the back-scattering of the gauge fields which can induce inhomogeneities in the axion field. A recent study suggests that including this effect could make the suppression of the axion relic abundance less efficient \cite{Kitajima:2017peg}. This will not have a dramatic impact on the GW signal, since it is dominantly produced during the large initial drop in the axion energy density. However, the parameter space which is consistent with cosmology would be reduced unless additional mechanisms were invoked to suppress the axion relic abundance.

\subsection{Gravitational Waves from Relaxation with Particle Production?}
Tachyonic particle production can also be used as an alternative to Hubble friction in the relaxion mechanism \cite{Hook:2016mqo}. 
As explored in detail in Ref.~\cite{Fonseca:2018xzp}, when relaxation with particle production takes place after inflation, the maximum allowed cutoff is around  $\Lambda \sim 10^{5}-10^{6}$ GeV. Assuming the relaxion dominates the energy density of the universe at the time of particle production such that the energy budget available for gravitational waves is maximized, the Hubble parameter at the time of emission is $H_{*} \sim \Lambda^{2}/M_{P}$. The most negative tachyonic frequency in the case of the relaxion is $k_{\rm peak} \sim v_{\rm EW}$ by construction, where $v_{\rm EW} = 246$ GeV is the electroweak scale. We then estimate the peak of the gravitational wave spectrum at the present time using Eqs.~\ref{eq:scal_rel} and~\ref{eq:amp_RS_simple} and find
\begin{equation}
\begin{split}
\Omega_{\rm GW}^{0} &\sim 1.67\times10^{-4} \, g_{\rho,*}^{-1/3}  \left(\frac{\Lambda^2}{v_{\rm EW}\, M_P } \right)^2 \\
&\sim 10^{-26}-10^{-22} \,,
\end{split}
\end{equation}
where we have used $ g_{\rho,*} = g_{s,*} = 106.8$ as the number of relativistic degrees of freedom at the time of emission and the range in the result corresponds to considering a maximum cutoff of $\Lambda = 10^{5} - 10^{6}$~GeV. The peak frequency at the time of emission is given by $v_{\rm EW}$ and redshifting to the present time yields
\begin{equation}
 f_{0} \approx \left(\frac{g_{s,{\rm eq}}}{g_{s,*}}\right)^{\frac{1}{3}} \left(\frac{T_{\rm 0}}{\Lambda}\right) v_{\rm EW} \sim 10^{6}-10^{7} \,\, {\rm Hz} \,.
\end{equation}
This result, while simply an order of magnitude estimation, illustrates why gravitational radiation from relaxation with particle production would be very challenging to detect. The present time peak frequency is far above the reach of present or planned future detectors, and while the spectrum has a tail which extends to lower frequencies, even the peak amplitude is many orders of magnitude below the sensitivity of any planned future experiment.

\section{Conclusions}
\label{s:conclusions}
We propose a novel method to search for axions or ALPs that couple to dark photons, without relying on SM couplings.
The essential dynamics occur in a radiation dominated era, which is distinct from the inflation and preheating scenarios explored previously in the literature. If the axion-dark photon coupling is sizeable, a tachyonic instability followed by a phase of parametric resonance occurs as the axion rolls, allowing the energy of the axion to be efficiently transferred to dark photons.
This suppression of the axion relic abundance allows for larger decay constants without tuning the initial misalignment angle. In this context, it would be important to to understand the effects of the gauge field back-scattering, neglected in our simulation.

The same mechanism causes vacuum fluctuations of the dark photon to experience exponential growth, resulting in time-dependent anisotropies that source gravitational waves.  The GW amplitude is controlled mainly by $f/M_{P}$ and the frequency by the ALP mass, allowing for a wide region of parameter space to be explored by future GW experiments as we show with our numerical results for the GW spectra in Figure~\ref{fig:money_plot}. In addition to the distinct shape of the spectrum, its chiral nature could help distinguish it from other cosmological sources of stochastic GW backgrounds. 

Complementary to GW searches, the region of parameter space which gives an observable GW signal will also be probed with future CMB experiments which will constrain the production of dark radiation. It is exciting that these astrophysical observations provide a new window to probe invisible axion models.

\subsection*{Acknowledgments}
We thank Gustavo Marques-Tavares, Nayara Fonseca, Enrico Morgante, Toby Opferkuch, Joachim Kopp, Will Shepherd, Valerie Domcke, and Alexander Westphal for useful discussions. The work of CSM was supported
by the Alexander von Humboldt Foundation, in the framework of the Sofja Kovalevskaja Award 2016, endowed by the German Federal Ministry of Education and Research. Our work is also supported by the DFG Cluster of Excellence PRISMA (EXC 1098).


%

  
\newpage
\onecolumngrid
\appendix
\section{Parametric Resonance}
\label{appB}
Modes which leave the tachyonic production band can still experience significant growth due to parametric resonance with the coherently oscillating axion field, which allows for additional suppression of the relic abundance of the axion.
In order to get a better understanding of the interplay between the tachyonic and the parametric resonance bands, we write Eq.~\ref{eq:mode_eq} in the form of the Mathieu equation using Eq.~\ref{eq:phi_t} (e.g. \cite{Kofman:1997yn,Dufaux:2006ee,Adshead:2015pva}) 
\begin{equation}
\frac{d^{2} v_{\pm} (k,z)}{d z^2} + \left[A_k \pm 2qF(z)\right]v_{\pm} (k,z)=0\,,
\label{eq:mathieu_eq}
\end{equation}
where $z=mt/2$, $F(z)$ is a harmonic function with unit amplitude, and we have
\begin{equation}
A_k \equiv 4\left(\frac{k}{a m}\right)^2, \qquad q\equiv\pm \, 2\frac{\alpha}{f} \left(\frac{k}{am}\right)\left(\frac{a_{\rm osc}}{a}\right)^{3/2}\phi_{\rm osc}\,.
\label{eq:Ak_and_q}
\end{equation}
While the backreaction from dark photon production is small, the tachyonic instabilities and parametric resonance can be studied using the stability and instability regions in the $(A_k,q)$ plane. The tachyonic regime is set by $A_k -2q<0$, or for comoving momenta $k < \theta\alpha am (a_{\rm osc}/a)^{3/2}$, in agreement with the discussion in Section~\ref{s:darkphoton}. Roughly speaking, the broad resonance regime is given by $q \gg 1$ while the narrow resonance regime occurs for $q \lesssim 1$. The broad resonance regime is given initially by $k \gg ma_{\rm osc}/(2\alpha\theta)$ and  $k\gg m a_{\rm osc} (\alpha\theta/16)^{2/3}$ when the tachyonic band closes. One can show that for $\alpha\theta \gg 1$, the broad resonance regime includes entire tachyonic production band throughout its evolution. Thus, we expect that the initial tachyonic growth phase also includes effects from broad parametric resonance which cannot be easily decoupled. 
\begin{figure}[h!]
\centering 
\includegraphics[scale=0.85]{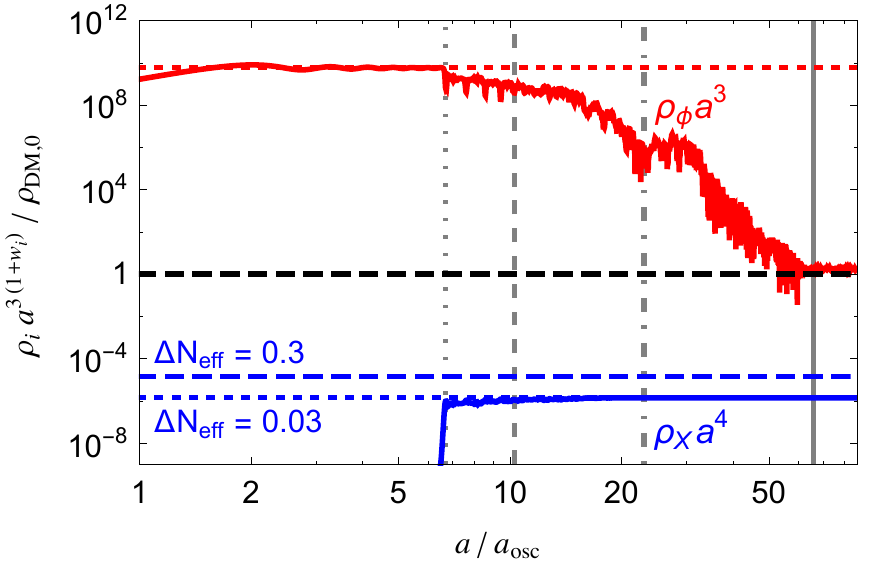} \hspace{5mm}
\includegraphics[scale=0.85]{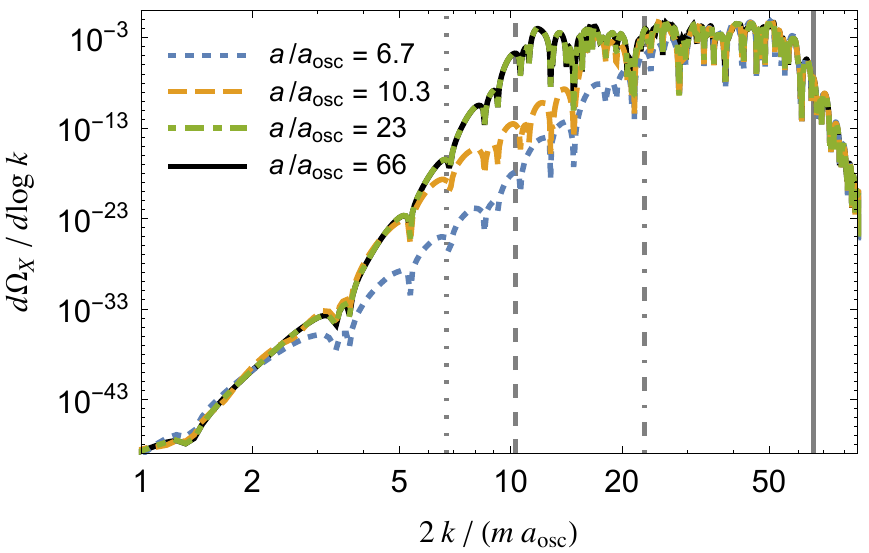}
\caption{Left: Comoving energy densities normalized to the present time dark matter energy density with the scale factor normalized to unity at the present time. Right: Dark photon power spectrum where the vertical gray lines show the position of the narrow resonance band $k = ma/2$ for four different values of the scale factor.  The line styles correspond to: Dotted: Initial large energy drop, Dashed: Tachyonic band closure as approximated by Eq.~\ref{eq:TPB}, Dot Dashed: Narrow parametric resonance band enters the most pumped part of the gauge power spectrum, Solid: Narrow parametric resonance band leaves the most pumped part of the spectrum and energy transfer from the axion to dark photons ends when the scale factor is $\approx\alpha\theta$. The plots here use the ALP 2 model parameters as shown in Table~\ref{tab:GW_spec_params}.}
\label{fig:CME_and_GPS2}
\end{figure}

Once friction from dark photon production becomes important, we expect that the corresponding rapid drop in the axion amplitude (which is not captured by Eq.~\ref{eq:Ak_and_q}) will cause a transition from broad to narrow parametric resonance. The narrow resonance bands are given by $A_k\approx n^2$ ($n=1,2,...$), so the lowest $k$ band corresponds to $n=1$ or $k=ma/2$. The study in Ref.~\cite{Adshead:2015pva} finds that once $\alpha\phi/f \lesssim 1$, then the narrow resonance band around $k=ma/2$ is the most important and that the enhancement from narrow parametric resonance falls off sharply for $k<ma/2$ since there are no more narrow bands for lower $k$. This implies that the energy transfer from the axion to dark photons should end when the largest comoving scale which experienced significant tachyonic growth becomes less than $ma/2$. Estimating this scale as the initial value of $\tilde{k}$ leads to the condition
\begin{equation}
a_{f}/a_{\rm osc} \approx \alpha\theta \,
\end{equation}
where $a_{f}$ is the scale factor when energy transfer from the axion to dark photons ends. We can also give a qualitative explanation of the features observed in the evolution of the axion comoving energy density as being due to the narrow parametric resonance band at $k=ma/2$ entering and then later  exiting the most highly pumped part of the gauge power spectrum. This is shown more explicitly in Figure~\ref{fig:CME_and_GPS2}.

\begin{figure}
\centering
\includegraphics[scale=1.0]{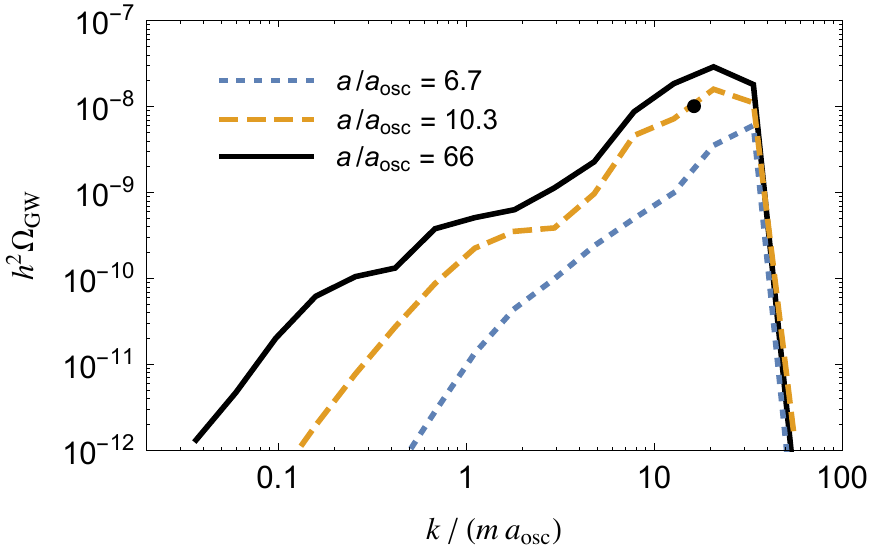}
\caption{Contribution to the GW spectrum at different times as parametrized by the scale factor. The blue dotted line corresponds to the  large initial drop in the axion's energy. The yellow dashed line gives the contribution up to the time where the tachyonic band is estimated to close, in good agreement with our analytic estimate from section~\ref{s:gws}. The solid black line gives the total spectrum after energy transfer has ended. The line styles and colors match the definitions in Figure~\ref{fig:CME_and_GPS2}. The plot uses the ALP 2 model parameters and the GW spectra were computed with $N=50$ modes.}
\label{fig:tach_vs_tot_GW}
\end{figure}

While the parametric resonance phase is essential for suppression of the axion relic abundance, the GW spectrum is dominantly determined by the initial tachyonic phase, where a large fraction of the axion's initial energy is transferred to radiation. This can be seen in Figure~\ref{fig:tach_vs_tot_GW}, where we show the GW spectrum at different times, including the large initial drop in the axion's energy, the closing of the tachyonic band, and the total spectrum when energy transfer has ended which includes the contribution from the narrow parametric resonance.

\section{Computation of the Gravitational Wave Spectrum}
\label{app:first_app}
The energy momentum tensor and perturbed metric are
\begin{equation}
T_{\mu\nu} = - X_{\mu}^{\,\,\alpha}X_{\nu\alpha} - \mathcal{L} g_{\mu\nu}\,, \hspace{20mm} ds^{2} = a(\tau)^{2} \left[ d\tau^{2} - (\delta_{ij}+h_{ij}) dx^{i}dx^{j}\right]  \,,
\end{equation}
and the Einstein equations give the following wave equation for $h_{ij}$
\begin{equation}
h''_{ij}({\bf x},\tau) + 2aH h'_{ij}({\bf x},\tau) -\nabla^{2} h_{ij}({\bf x},\tau) = \frac{2}{M_{P}^{2}}\Pi_{ij}({\bf x},\tau) \, , 
\label{eomGW}
\end{equation}
where $M_{P} = (8\pi G)^{-1/2} = 2.44\times 10^{18}$ GeV, $\Pi_{ij}({\bf x},\tau) = \Lambda_{ij}^{kl}T_{kl}({\bf x},\tau)$ is the anisotropic stress energy, and $\Lambda_{ij}^{kl} = \Lambda_{i}^{k} \Lambda_{j}^{l} - \frac{1}{2}\Lambda_{ij}\Lambda^{kl}$ is the transverse traceless projector with $\Lambda_{ij} = \delta_{ij} -\partial_{i}\partial_{j}/\nabla^{2}$. One can easily see that the part of $T_{ij}$ which is proportional to $g_{ij} \propto \delta_{ij}$ will not source gravitational waves. Thus, the relevant part of the stress energy tensor is
\begin{equation}
\begin{split}
T_{ij} &= -X_{i}^{\,\,\alpha}X_{j\alpha} = -g^{00}X_{i0}X_{j0}  - g^{kl}X_{il}X_{jk} = -\frac{1}{a^{2}}\left(X_{0i}X_{0j}  + X_{ik}X_{kj}\right) = -\frac{1}{a^{2}}\left(E_{i}E_{j} + B_{i}B_{j}\right) \,,
\end{split}
\end{equation}
where we have used the result $X_{ik}X_{kj} = B_{i}B_{j} - B^{2}\delta_{ij}$ and thrown away the isotropic term. 
Defining $\bar{h}_{ij} = a h_{ij}$ and taking the Fourier transform of the GW equation, we find
\begin{equation}
\bar{h}''_{ij}({\bf k},\tau) +\left(k^{2} -\frac{a''}{a}\right) \bar{h}_{ij}({\bf k},\tau) = \frac{2a}{M_{P}^{2}}\Pi_{ij}({\bf k},\tau) \, , 
\end{equation}
and if we throw away the $a''$ term which vanishes in the radiation dominated era where $a\propto \tau$, the solution for $h_{ij}$ is
 \begin{equation}
h_{ij} ({\bf k},\tau)= \frac{2}{a(\tau) M_{P}^{2}} \int_{\tau_i}^{\tau} d\tau' a(\tau') \mathcal{G}(k,\tau,\tau') \Pi_{ij}({\bf k},\tau')  \, , \hspace{12mm}  \mathcal{G}(k,\tau,\tau') = \frac{1}{k}\sin\left[ k(\tau-\tau')\right] \, ,
\end{equation}
where $\mathcal{G}$ is the retarded (or causal) Green's function for the d'Alembert operator.
The gravitational wave power spectrum is given by
\begin{equation}
\frac{d\Omega_{\rm GW}}{d \log k} =\frac{1}{\rho_{\rm tot}} \frac{M_{P}^{2}k^{3}}{8\pi^{2}a^{2}} \mathcal{P}_{h'}({\bf k},\tau) \, ,
\end{equation}
with $ \langle h'_{ij}({\bf k},\tau)h^{'*}_{ij}({\bf k}',\tau) \rangle = (2\pi)^{3}\mathcal{P}_{h'}({\bf k},\tau) \delta({\bf k}-{\bf k}')$. Inserting the solution for $h_{ij}$, we find
\begin{equation}
\frac{d\Omega_{\rm GW}}{d \log k} =\frac{1}{\rho_{\rm tot}} \frac{k^{3}}{4\pi^{2}a^{4}M_{P}^{2}}\int_{\tau_i}^{\tau} d\tau' d\tau'' a(\tau')a(\tau'') \cos\left[k(\tau'-\tau'') \right]\Pi^{2}({\bf k},\tau',\tau'') \, 
\end{equation}
where we have averaged over one period $\Delta\tau = 2\pi/k$ which gives a factor of 1/2. The function $\Pi^{2}({\bf k},\tau',\tau'') $ is called the Unequal Time Correlator (UTC) and is defined via  $ \langle \Pi_{ij}({\bf k},\tau)\Pi^{*}_{ij}({\bf k}',\tau') \rangle = (2\pi)^{3}\Pi^{2}({\bf k},\tau,\tau') \delta({\bf k}-{\bf k}')$. We now turn to computing this object. 

\subsection{Unequal Time Correlator}
The Fourier transform of the anisotropic stress requires a convolution
\begin{equation}
\Pi_{ij}({\bf k},\tau) =-\frac{\Lambda_{ij}^{kl}({\bf k}) }{a^{2}} \int \frac{d^{3}q}{(2\pi)^3} \left[ E_{k}({\bf q})E_{l}({\bf k-q}) + B_{k}({\bf q})B_{l}({\bf k-q})   \right] \, .
\end{equation}
Working in the Coulomb gauge defined by $\vec{\nabla}\cdot\vec{X} =0$, we have $X_{0} = 0$. Taking the Fourier transform of $E$ and $B$ yields $E_{i}({\bf q}) = X'_{i}({\bf q}) = v'_{\lambda} ({\bf q}, \tau)  \veps_{i}^{\lambda}({\bf q})$. The transform for $B$ gives $B_{i}({\bf q}) = -i \epsilon_{ijk}q_{j}X_{k}({\bf q}) = \lambda \, q\, v_{\lambda} ({\bf q}, \tau)  \veps_{i}^{\lambda}({\bf q})$. Putting it all together and leaving the sum over helicities implied, we have
 \begin{equation}
\Pi_{ij}({\bf k},\tau) = \int \frac{d^{3}q}{(2\pi)^3} \Theta_{ij}^{\lambda_{1}\lambda_{2}}({\bf q}, {\bf k}) \, \mathcal{S}_{\lambda_{1}\lambda_{2}}({\bf q}, {\bf k}, \tau) \, ,
\end{equation}
with an angular function defined as
\begin{equation}
 \Theta_{ij}^{\lambda_{1}\lambda_{2}}({\bf q}, {\bf k}) \equiv \Lambda_{ij}^{kl}({\bf k})\veps_{k}^{\lambda_{1}}({\bf q}) \veps_{l}^{\lambda_{2}}({\bf k-q}) \, ,
\end{equation}
and a source function
\begin{equation}
\mathcal{S}_{\lambda_{1}\lambda_{2}}({\bf q}, {\bf k}, \tau) \equiv -\frac{1}{a^{2}} \bigg[\lambda_{1}\lambda_{2} \, |{\bf q}| |{\bf k}-{\bf q}|\,v_{\lambda_{1}}({\bf q})v_{\lambda_{2}}({\bf k-q}) + v'_{\lambda_{1}}({\bf q})v'_{\lambda_{2}}({\bf k-q}) \bigg] \,.
\end{equation}
We now promote $\Pi$ to an operator via $v_{\lambda}({\bf q}) \rightarrow \hat{a}_{\lambda}({\bf q})v_{\lambda}({\bf q})$, where the operators satisfy the commutation relation
\begin{equation}
[\hat{a}_{\lambda}({\bf q}),\hat{a}^{\dagger}_{\lambda'}({\bf q'})] = (2\pi)^{3} \delta_{\lambda\lambda'} \delta({\bf q-q}') \,.
\end{equation}
The object we need to compute is
\begin{equation}
\begin{split}
 &\langle 0|\Pi_{ij}({\bf k},\tau)\Pi^{*}_{ij}({\bf k}',\tau') |0\rangle =  \int \frac{d^{3}q}{(2\pi)^3} \int \frac{d^{3}q'}{(2\pi)^3} \Theta_{ij}^{\lambda_{1}\lambda_{2}}({\bf q}, {\bf k}) \Theta_{ij}^{*\lambda'_{1}\lambda'_{2}}({\bf q'}, {\bf k'}) \\
 &\times  \mathcal{S}_{\lambda_{1}\lambda_{2}}({\bf q}, {\bf k}, \tau) \mathcal{S}^{*}_{\lambda'_{1}\lambda'_{2}}({\bf q'}, {\bf k'}, \tau')\,   \langle 0| \hat{a}_{\lambda_{1}}({\bf q}) \hat{a}_{\lambda_{2}}({\bf k-q}) \hat{a}^{\dagger}_{\lambda'_{1}}({\bf q'}) \hat{a}^{\dagger}_{\lambda'_{2}}({\bf k'-q'}) |0\rangle \, .
 \end{split}
\end{equation}
Using the commutation relation, one can show that 
\begin{equation}
\begin{split}
 &\langle 0| \hat{a}_{\lambda_{1}}({\bf q}) \hat{a}_{\lambda_{2}}({\bf k-q}) \hat{a}^{\dagger}_{\lambda'_{1}}({\bf q'}) \hat{a}^{\dagger}_{\lambda'_{2}}({\bf k'-q'}) |0\rangle \\
  &=(2\pi)^6 \delta({\bf k-k}') \left[ \delta_{\lambda_{1}\lambda'_{1}}\delta_{\lambda_{2}\lambda'_{2}}\delta({\bf q-q'}) + \delta_{\lambda_{1}\lambda'_{2}}\delta_{\lambda_{2}\lambda'_{1}}\delta({\bf k-q-q'}) \right] \, .
 \end{split}
 \label{eq:expV}
\end{equation}
The first term matches the helicities and $q$ to $q'$ whereas the second term sends $q' \rightarrow k-q$ and exchanges the helicities. It is easy to see that $\mathcal{S}$ is invariant under a transformation of the second type, and because $\Lambda_{ij}^{kl} = \Lambda_{ij}^{lk}$, so is $\Theta$. Put explicitly, we have
\begin{equation}
\mathcal{S}_{\lambda_{2}\lambda_{1}}({\bf k-q}, {\bf k}, \tau)  = \mathcal{S}_{\lambda_{1}\lambda_{2}}({\bf q}, {\bf k}, \tau)\,, \hspace{13mm}
 \Theta_{ij}^{\lambda_{2}\lambda_{1}}({\bf k-q}, {\bf k}) =  \Theta_{ij}^{\lambda_{1}\lambda_{2}}({\bf q}, {\bf k}) \, ,
 \label{eq:syms}
\end{equation}
therefore we can just take twice the first term in Eq.~\ref{eq:expV} to arrive at the result
\begin{equation}
 \langle 0| \hat{a}_{\lambda_{1}}({\bf q}) \hat{a}_{\lambda_{2}}({\bf k-q}) \hat{a}^{\dagger}_{\lambda'_{1}}({\bf q'}) \hat{a}^{\dagger}_{\lambda'_{2}}({\bf k'-q'}) |0\rangle \\
  =2(2\pi)^6 \delta({\bf k-k}')\delta_{\lambda_{1}\lambda'_{1}}\delta_{\lambda_{2}\lambda'_{2}}\delta({\bf q-q'}) \, .
\end{equation}
Comparing to the definition $ \langle \Pi_{ij}({\bf k},\tau)\Pi^{*}_{ij}({\bf k}',\tau') \rangle = (2\pi)^{3}\Pi^{2}({\bf k},\tau,\tau') \delta({\bf k}-{\bf k}')$, we find for the UTC
\begin{equation}
\Pi^{2}({\bf k},\tau,\tau') = 2 \sum_{\lambda_{1},\lambda_{2} = \pm}\int \frac{d^{3}q}{(2\pi)^{3}} |\Theta_{\lambda_{1}\lambda_{2}}({\bf k-q}, {\bf k})|^{2} \mathcal{S}_{\lambda_{1}\lambda_{2}}({\bf q}, {\bf k}, \tau) \mathcal{S}^{*}_{\lambda_{1}\lambda_{2}}({\bf q}, {\bf k}, \tau')\,  \, ,
\end{equation}
with
\begin{equation}
\begin{split}
|\Theta_{\lambda_{1}\lambda_{2}}({\bf k-q}, {\bf k})|^{2}
&\equiv  \Lambda_{ij}^{ab}({\bf k})\Lambda_{ij}^{cd}({\bf k}) \, \veps_{a}^{\lambda_{1}}({\bf q}) \veps_{b}^{\lambda_{2}}({\bf k-q}) \veps_{c}^{-\lambda_{1}}({\bf q}) \veps_{d}^{-\lambda_{2}}({\bf k-q})\\
&= \veps_{a}^{-\lambda_{1}}({\bf q}) \veps_{b}^{-\lambda_{2}}({\bf k-q})\  \Lambda^{abcd}({\bf k})\  \veps_{c}^{\lambda_{1}}({\bf q}) \veps_{d}^{\lambda_{2}}({\bf k-q})\, .
\end{split}
\end{equation}
\subsection{Polarization Vectors and Angular Function}
Using  $\Lambda^{abcd} = \Lambda^{ac}\Lambda^{bd}-\frac{1}{2}\Lambda^{ab}\Lambda^{cd}$ where the individual projectors can be written in terms of the polarization vectors as
\begin{equation}
\Lambda_{ij}({\bf k}) =  \veps_{i}^{+}({\bf k})\veps_{j}^{-}({\bf k}) + \veps_{i}^{-}({\bf k})\veps_{j}^{+}({\bf k}) \, ,
\label{eq:proj}
\end{equation}
and keeping in mind that the projection operator is only acting on symmetric tensors we can write
\begin{equation}
 \Lambda_{abcd} = \veps_{a}^{+}({\bf k})\veps_{b}^{+}({\bf k})\ \veps_{c}^{-}({\bf k})\veps_{d}^{-}({\bf k})+ \veps_{a}^{-}({\bf k})\veps_{b}^{-}({\bf k})\ \veps_{c}^{+}({\bf k})\veps_{d}^{+}({\bf k})\, .
\end{equation}
We notice that the term $\veps_{a}^{\lambda}({\bf k})\veps_{b}^{\lambda}({\bf k})$ picks up a phase $\exp(i2\lambda\phi)$ under a rotation about ${\bf k}$ by an angle $\phi$, so we identify the two summands in the equation above as helicity projectors that leave us with the part of the anisotropic stress sourcing one particular GW circular polarization. Since the wave equation for $h_{ij}$ is linear, the different polarizations do not interfere and it is sufficient to introduce 
\begin{equation}
 |\Theta^{\lambda}_{\lambda_{1}\lambda_{2}}({\bf k-q}, {\bf k})|^{2}
 \equiv \veps_{a}^{-\lambda_{1}}({\bf q}) \veps_{b}^{-\lambda_{2}}({\bf k-q})\ 
 \left[\veps_{a}^{\lambda}({\bf k})\veps_{b}^{\lambda}({\bf k})\ \veps_{c}^{-\lambda}({\bf k})\veps_{d}^{-\lambda}({\bf k})\right]\  
 \veps_{c}^{\lambda_{1}}({\bf q}) \veps_{d}^{\lambda_{2}}({\bf k-q})\,
\end{equation}
in order to study the polarization of the gravitational wave spectrum. We use the relation
\begin{equation}
|\veps_{i}^{\lambda_{1}}({\bf k})\veps_{i}^{\lambda_{2}}({\bf q})|^{2} = \frac{1}{4} \left(1 - \lambda_{1}\lambda_{2} \, \frac{{\bf k} \cdot {\bf q}}{|{\bf k}||{\bf q}|} \right)^{2} \, ,
\end{equation}
to arrive at
\begin{equation}
\begin{split}
&|\Theta^\lambda_{\lambda_{1}\lambda_{2}}(k,q,\theta)|^{2} \equiv |\Theta^\lambda_{\lambda_{1}\lambda_{2}}({\bf q}, {\bf k})|^2 =\frac{1}{16} (1+\lambda\lambda_{1}\Phi_{1})^2 (1+\lambda\lambda_{2} \Phi_{2})^2 \\
&|\Theta_{\lambda_{1}\lambda_{2}}(k,q,\theta)|^{2} \equiv |\Theta_{\lambda_{1}\lambda_{2}}({\bf q}, {\bf k})|^2 = |\Theta^+_{\lambda_{1}\lambda_{2}}(k,q,\theta)|^{2}+|\Theta^-_{\lambda_{1}\lambda_{2}}(k,q,\theta)|^{2}\, ,
\label{ThetaCalc}
\end{split}
\end{equation}
with the functions $\Phi_{1}\equiv \Phi_{1}(k,q,\theta) $ and $\Phi_{2}\equiv \Phi_{2}(k,q,\theta)$ defined as
\begin{equation}
\Phi_{1}(k,q,\theta) \equiv \frac{{\bf k} \cdot {\bf q}}{|{\bf k}||{\bf q}|} = \cos\theta \,, \hspace{10mm} \Phi_{2}(k,q,\theta) \equiv \frac{{\bf k} \cdot {\bf (k-q)}}{|{\bf k}||{\bf k-q}|} = \frac{k -q\cos\theta}{\sqrt{k^{2}+q^{2}-2qk\cos\theta}}  \, ,
\end{equation}
where $\theta$ is the angle between ${\bf k}$ and ${\bf q}$. The function $|\Theta|^{2}$ has a few nice properties which are worth pointing out. The first is an exchange symmetry ($1\leftrightarrow 2$) under which it is invariant. Perhaps unsurprisingly, the transformation $\Phi_{1} \rightarrow \Phi_{2}$ is equivalent to ${\bf q} \rightarrow {\bf k-q}$, so the symmetry we identified in Eq.~\ref{eq:syms} for $\Theta$ has been preserved. Additionally, we see that $|\Theta^\lambda_{\lambda_{1}\lambda_{2}}|^{2}$ is invariant under $\lambda_{1}\rightarrow -\lambda_{1}$, $\lambda_{2} \rightarrow -\lambda_{2}$ and $\lambda\rightarrow -\lambda$. The domain of both $\Phi_{1}$ and $\Phi_{2}$ is $[-1,1]$, which one can use to check that the range of $|\Theta|^{2}$ is $[0,1]$, thus the function is positive definite and unitary.

\subsection{Gravitational Wave Power Spectrum: Result}
Using the identity $\cos[k(x-y)] = \cos(kx)\cos(ky) + \sin(kx)\sin(ky)$, the (polarized) GW power spectrum can now be written as
\begin{equation}
\begin{split}
&\frac{d\Omega^{(\lambda)}_{\rm GW}}{d \log k} =\frac{1}{\rho_{\rm tot}} \frac{k^{3}}{2\pi^{2}a^{4}M_{P}^{2}} \sum_{\lambda_{1},\lambda_{2} = \pm}\int \frac{d^{3}q}{(2\pi)^{3}} |\Theta^{(\lambda)}_{\lambda_{1}\lambda_{2}}(k,q,\theta)|^{2} \\
& \times  \int_{\tau_i}^{\tau} d\tau' d\tau'' a(\tau')a(\tau'') \left[\cos(k\tau')\cos(k\tau'')+\sin(k\tau')\sin(k\tau'')\right] \mathcal{S}_{\lambda_{1}\lambda_{2}}({\bf q}, {\bf k}, \tau') \mathcal{S}^{*}_{\lambda_{1}\lambda_{2}}({\bf q}, {\bf k}, \tau'')\, ,
\end{split}
\end{equation}
and we see that the integrals factorize, so we define
\begin{equation}
I_{c}^{\lambda_{1}\lambda_{2}} (k,q,\theta,\tau) \equiv  \int_{\tau_i}^{\tau} d\tau' a(\tau') \cos(k\tau') \, \mathcal{S}_{\lambda_{1}\lambda_{2}}({\bf q}, {\bf k}, \tau') \, ,
\end{equation}
\begin{equation}
I_{s}^{\lambda_{1}\lambda_{2}} (k,q,\theta,\tau) \equiv  \int_{\tau_i}^{\tau} d\tau' a(\tau') \sin(k\tau') \, \mathcal{S}_{\lambda_{1}\lambda_{2}}({\bf q}, {\bf k}, \tau') \, ,
\end{equation}
such that the equation for the GW power spectrum takes the form
\begin{equation}
\begin{split}
&\frac{d\Omega^{(\lambda)}_{\rm GW}}{d \log k} =\frac{1}{\rho_{\rm tot}} \frac{k^{3}}{2\pi^{2}a^{4}M_{P}^{2}}  \sum_{\lambda_{1},\lambda_{2} = \pm}\int \frac{d^{3}q}{(2\pi)^{3}} |\Theta^{(\lambda)}_{\lambda_{1}\lambda_{2}}(k,q,\theta) |^{2}\left(|I_{c}^{\lambda_{1}\lambda_{2}} (k,q,\theta,\tau)|^{2} + |I_{s}^{\lambda_{1}\lambda_{2}} (k,q,\theta,\tau)|^{2}\right) \, .
\end{split}
\end{equation}
Working in spherical coordinates $(q,\phi,\theta)$ where $\phi$ is the azimuthal angle and $\theta$ the polar angle, we are free to choose ${\bf k} = k\, (0,0,1)$ and ${\bf q} = q \, (\sin\theta \cos\phi, \sin\theta\sin\phi, \cos\theta)$. Performing the $\phi$ integration and changing variables $x=\cos\theta$, we have
\begin{equation}
\begin{split}
&\frac{d\Omega^{(\lambda)}_{\rm GW}}{d \log k} =\frac{1}{\rho_{\rm tot}} \frac{k^{3}}{8\pi^{4} a^{4}M_{P}^{2}}  \sum_{\lambda_{1},\lambda_{2} = \pm} \int_{0}^{\infty} q^{2}dq \int_{-1}^{1} dx\, \,  |\Theta^{(\lambda)}_{\lambda_{1}\lambda_{2}}(k,q,x)|^{2} \left(|I_{c}^{\lambda_{1}\lambda_{2}} (k,q,x,\tau)|^{2} + |I_{s}^{\lambda_{1}\lambda_{2}} (k,q,x,\tau)|^{2}\right) \, .
\end{split}
\end{equation}
 Because many terms depend on $|{\bf k}-{\bf q}|$, it is natural to trade the $x$ integration for integration over $l \equiv |{\bf k}-{\bf q}| = \sqrt{k^{2}+q^{2}-2kq x}$
\begin{equation}
\begin{split}
&\frac{d\Omega^{(\lambda)}_{\rm GW}}{d \log k} =\frac{1}{\rho_{\rm tot}} \frac{k^{2}}{8\pi^{4} a^{4}M_{P}^{2}}  \sum_{\lambda_{1},\lambda_{2} = \pm} \int_{0}^{\infty} dq \, q \int_{|k-q|}^{|k+q|} dl \, l \,\, | \Theta^{(\lambda)}_{\lambda_{1}\lambda_{2}}(k,q,l)|^{2} \left(|I_{c}^{\lambda_{1}\lambda_{2}} (k,q,l,\tau)|^{2} + |I_{s}^{\lambda_{1}\lambda_{2}} (k,q,l,\tau)|^{2}\right) .
\end{split}
\end{equation}

\subsection{Numerics}
In the case where one vector helicity dominates (taken to be ``+" ), we have
\begin{equation}
\begin{split}
&\frac{d\Omega^{(\lambda)}_{\rm GW}}{d \log k} =\frac{1}{\rho_{\rm tot}} \frac{k^{2}}{8\pi^{4} a^{4}M_{P}^{2}}  \int_{0}^{\infty} dq \, q \int_{|k-q|}^{|k+q|} dl \, l \,\,  |\Theta^{(\lambda)}_{++}(k,q,l)|^{2} \left(|I_{c}^{++} (k,q,l,\tau)|^{2} + |I_{s}^{++} (k,q,l,\tau)|^{2}\right) \, .
\end{split}
\end{equation}
Of use will be
\begin{equation}
I_{c}^{++} (k,q, l,\tau) =- \int_{\tau_i}^{\tau} \frac{d\tau'}{a(\tau')} \cos(k\tau') \bigg[q l\, v_{+}(q,\tau')v_{+}(l,\tau') + v'_{+}(q,\tau')v'_{+}(l,\tau') \bigg]  \, ,
\end{equation}
\begin{equation}
I_{s}^{++} (k,q, l,\tau) = -\int_{\tau_i}^{\tau} \frac{d\tau'}{a(\tau')} \sin(k\tau') \bigg[ q l\,  v_{+}(q,\tau')v_{+}(l,\tau') + v'_{+}(q,\tau')v'_{+}(l,\tau') \bigg]  \, .
\end{equation}
For evaluating the angular function in terms of $l$, we have
\begin{equation}
\Phi_{1}(k,q,l) = \frac{k^{2} + q^{2} - l^{2}}{2kq}  \,, \hspace{15mm} \Phi_{2}(k,q,l) = \frac{k^{2} - q^{2} + l^{2}}{2kl}   \, ,
\end{equation}
which transform into each other under the exchange $q \leftrightarrow l$ as expected and together with \eq{ThetaCalc} allow a straight forward computation of $|\Theta|^2$.
We discretize the $q$ and $l$ integrations via the replacement
\begin{equation}
\int_{0}^{\infty} dq \, q \int_{|k-q|}^{|k+q|} dl \, l  \hspace{5mm} \longrightarrow  \hspace{5mm}  \sum_{q=0}^{q_{\rm max}}\Delta q \, q \, \sum_{l \, \in \, L} \Delta l \, l \, ,
\end{equation}
where $q_{\rm max} = \theta \alpha m a_{\rm osc}$ and $L = L(k,q)$ is a list of simulated modes between $|k-q|$ and $|k+q|$.


\begin{thebibliography}{42}%
\makeatletter
\providecommand \@ifxundefined [1]{%
 \@ifx{#1\undefined}
}%
\providecommand \@ifnum [1]{%
 \ifnum #1\expandafter \@firstoftwo
 \else \expandafter \@secondoftwo
 \fi
}%
\providecommand \@ifx [1]{%
 \ifx #1\expandafter \@firstoftwo
 \else \expandafter \@secondoftwo
 \fi
}%
\providecommand \natexlab [1]{#1}%
\providecommand \enquote  [1]{``#1''}%
\providecommand \bibnamefont  [1]{#1}%
\providecommand \bibfnamefont [1]{#1}%
\providecommand \citenamefont [1]{#1}%
\providecommand \href@noop [0]{\@secondoftwo}%
\providecommand \href [0]{\begingroup \@sanitize@url \@href}%
\providecommand \@href[1]{\@@startlink{#1}\@@href}%
\providecommand \@@href[1]{\endgroup#1\@@endlink}%
\providecommand \@sanitize@url [0]{\catcode `\\12\catcode `\$12\catcode
  `\&12\catcode `\#12\catcode `\^12\catcode `\_12\catcode `\%12\relax}%
\providecommand \@@startlink[1]{}%
\providecommand \@@endlink[0]{}%
\providecommand \url  [0]{\begingroup\@sanitize@url \@url }%
\providecommand \@url [1]{\endgroup\@href {#1}{\urlprefix }}%
\providecommand \urlprefix  [0]{URL }%
\providecommand \Eprint [0]{\href }%
\providecommand \doibase [0]{http://dx.doi.org/}%
\providecommand \selectlanguage [0]{\@gobble}%
\providecommand \bibinfo  [0]{\@secondoftwo}%
\providecommand \bibfield  [0]{\@secondoftwo}%
\providecommand \translation [1]{[#1]}%
\providecommand \BibitemOpen [0]{}%
\providecommand \bibitemStop [0]{}%
\providecommand \bibitemNoStop [0]{.\EOS\space}%
\providecommand \EOS [0]{\spacefactor3000\relax}%
\providecommand \BibitemShut  [1]{\csname bibitem#1\endcsname}%
\let\auto@bib@innerbib\@empty
\bibitem [{\citenamefont {Peccei}\ and\ \citenamefont
  {Quinn}(1977{\natexlab{a}})}]{Peccei:1977ur}%
  \BibitemOpen
  \bibfield  {author} {\bibinfo {author} {\bibfnamefont {R.~D.}\ \bibnamefont
  {Peccei}}\ and\ \bibinfo {author} {\bibfnamefont {H.~R.}\ \bibnamefont
  {Quinn}},\ }\href {\doibase 10.1103/PhysRevD.16.1791} {\bibfield  {journal}
  {\bibinfo  {journal} {Phys. Rev.}\ }\textbf {\bibinfo {volume} {D16}},\
  \bibinfo {pages} {1791} (\bibinfo {year} {1977}{\natexlab{a}})}\BibitemShut
  {NoStop}%
\bibitem [{\citenamefont {Peccei}\ and\ \citenamefont
  {Quinn}(1977{\natexlab{b}})}]{Peccei:1977hh}%
  \BibitemOpen
  \bibfield  {author} {\bibinfo {author} {\bibfnamefont {R.~D.}\ \bibnamefont
  {Peccei}}\ and\ \bibinfo {author} {\bibfnamefont {H.~R.}\ \bibnamefont
  {Quinn}},\ }\href {\doibase 10.1103/PhysRevLett.38.1440} {\bibfield
  {journal} {\bibinfo  {journal} {Phys. Rev. Lett.}\ }\textbf {\bibinfo
  {volume} {38}},\ \bibinfo {pages} {1440} (\bibinfo {year}
  {1977}{\natexlab{b}})},\ \bibinfo {note} {[,328(1977)]}\BibitemShut {NoStop}%
\bibitem [{\citenamefont {Svrcek}\ and\ \citenamefont
  {Witten}(2006)}]{Svrcek:2006yi}%
  \BibitemOpen
  \bibfield  {author} {\bibinfo {author} {\bibfnamefont {P.}~\bibnamefont
  {Svrcek}}\ and\ \bibinfo {author} {\bibfnamefont {E.}~\bibnamefont
  {Witten}},\ }\href {\doibase 10.1088/1126-6708/2006/06/051} {\bibfield
  {journal} {\bibinfo  {journal} {JHEP}\ }\textbf {\bibinfo {volume} {06}},\
  \bibinfo {pages} {051} (\bibinfo {year} {2006})},\ \Eprint
  {http://arxiv.org/abs/hep-th/0605206} {arXiv:hep-th/0605206 [hep-th]}
  \BibitemShut {NoStop}%
\bibitem [{\citenamefont {Arvanitaki}\ \emph {et~al.}(2010)\citenamefont
  {Arvanitaki}, \citenamefont {Dimopoulos}, \citenamefont {Dubovsky},
  \citenamefont {Kaloper},\ and\ \citenamefont
  {March-Russell}}]{Arvanitaki:2009fg}%
  \BibitemOpen
  \bibfield  {author} {\bibinfo {author} {\bibfnamefont {A.}~\bibnamefont
  {Arvanitaki}}, \bibinfo {author} {\bibfnamefont {S.}~\bibnamefont
  {Dimopoulos}}, \bibinfo {author} {\bibfnamefont {S.}~\bibnamefont
  {Dubovsky}}, \bibinfo {author} {\bibfnamefont {N.}~\bibnamefont {Kaloper}}, \
  and\ \bibinfo {author} {\bibfnamefont {J.}~\bibnamefont {March-Russell}},\
  }\href {\doibase 10.1103/PhysRevD.81.123530} {\bibfield  {journal} {\bibinfo
  {journal} {Phys. Rev.}\ }\textbf {\bibinfo {volume} {D81}},\ \bibinfo {pages}
  {123530} (\bibinfo {year} {2010})},\ \Eprint {http://arxiv.org/abs/0905.4720}
  {arXiv:0905.4720 [hep-th]} \BibitemShut {NoStop}%
\bibitem [{\citenamefont {Freese}\ \emph {et~al.}(1990)\citenamefont {Freese},
  \citenamefont {Frieman},\ and\ \citenamefont {Olinto}}]{Freese:1990rb}%
  \BibitemOpen
  \bibfield  {author} {\bibinfo {author} {\bibfnamefont {K.}~\bibnamefont
  {Freese}}, \bibinfo {author} {\bibfnamefont {J.~A.}\ \bibnamefont {Frieman}},
  \ and\ \bibinfo {author} {\bibfnamefont {A.~V.}\ \bibnamefont {Olinto}},\
  }\href {\doibase 10.1103/PhysRevLett.65.3233} {\bibfield  {journal} {\bibinfo
   {journal} {Phys. Rev. Lett.}\ }\textbf {\bibinfo {volume} {65}},\ \bibinfo
  {pages} {3233} (\bibinfo {year} {1990})}\BibitemShut {NoStop}%
\bibitem [{\citenamefont {Abbott}\ and\ \citenamefont
  {Sikivie}(1983)}]{Abbott:1982af}%
  \BibitemOpen
  \bibfield  {author} {\bibinfo {author} {\bibfnamefont {L.~F.}\ \bibnamefont
  {Abbott}}\ and\ \bibinfo {author} {\bibfnamefont {P.}~\bibnamefont
  {Sikivie}},\ }\href {\doibase 10.1016/0370-2693(83)90638-X} {\bibfield
  {journal} {\bibinfo  {journal} {Phys. Lett.}\ }\textbf {\bibinfo {volume}
  {B120}},\ \bibinfo {pages} {133} (\bibinfo {year} {1983})},\ \bibinfo {note}
  {[,URL(1982)]}\BibitemShut {NoStop}%
\bibitem [{\citenamefont {Dine}\ and\ \citenamefont
  {Fischler}(1983)}]{Dine:1982ah}%
  \BibitemOpen
  \bibfield  {author} {\bibinfo {author} {\bibfnamefont {M.}~\bibnamefont
  {Dine}}\ and\ \bibinfo {author} {\bibfnamefont {W.}~\bibnamefont
  {Fischler}},\ }\href {\doibase 10.1016/0370-2693(83)90639-1} {\bibfield
  {journal} {\bibinfo  {journal} {Phys. Lett.}\ }\textbf {\bibinfo {volume}
  {B120}},\ \bibinfo {pages} {137} (\bibinfo {year} {1983})},\ \bibinfo {note}
  {[,URL(1982)]}\BibitemShut {NoStop}%
\bibitem [{\citenamefont {Preskill}\ \emph {et~al.}(1983)\citenamefont
  {Preskill}, \citenamefont {Wise},\ and\ \citenamefont
  {Wilczek}}]{Preskill:1982cy}%
  \BibitemOpen
  \bibfield  {author} {\bibinfo {author} {\bibfnamefont {J.}~\bibnamefont
  {Preskill}}, \bibinfo {author} {\bibfnamefont {M.~B.}\ \bibnamefont {Wise}},
  \ and\ \bibinfo {author} {\bibfnamefont {F.}~\bibnamefont {Wilczek}},\ }\href
  {\doibase 10.1016/0370-2693(83)90637-8} {\bibfield  {journal} {\bibinfo
  {journal} {Phys. Lett.}\ }\textbf {\bibinfo {volume} {B120}},\ \bibinfo
  {pages} {127} (\bibinfo {year} {1983})},\ \bibinfo {note}
  {[,URL(1982)]}\BibitemShut {NoStop}%
\bibitem [{\citenamefont {Graham}\ \emph {et~al.}(2015)\citenamefont {Graham},
  \citenamefont {Kaplan},\ and\ \citenamefont {Rajendran}}]{Graham:2015cka}%
  \BibitemOpen
  \bibfield  {author} {\bibinfo {author} {\bibfnamefont {P.~W.}\ \bibnamefont
  {Graham}}, \bibinfo {author} {\bibfnamefont {D.~E.}\ \bibnamefont {Kaplan}},
  \ and\ \bibinfo {author} {\bibfnamefont {S.}~\bibnamefont {Rajendran}},\
  }\href {\doibase 10.1103/PhysRevLett.115.221801} {\bibfield  {journal}
  {\bibinfo  {journal} {Phys. Rev. Lett.}\ }\textbf {\bibinfo {volume} {115}},\
  \bibinfo {pages} {221801} (\bibinfo {year} {2015})},\ \Eprint
  {http://arxiv.org/abs/1504.07551} {arXiv:1504.07551 [hep-ph]} \BibitemShut
  {NoStop}%
\bibitem [{\citenamefont {Cardoso}\ \emph {et~al.}(2018)\citenamefont
  {Cardoso}, \citenamefont {Dias}, \citenamefont {Hartnett}, \citenamefont
  {Middleton}, \citenamefont {Pani},\ and\ \citenamefont
  {Santos}}]{Cardoso:2018tly}%
  \BibitemOpen
  \bibfield  {author} {\bibinfo {author} {\bibfnamefont {V.}~\bibnamefont
  {Cardoso}}, \bibinfo {author} {\bibfnamefont {O.~J.~C.}\ \bibnamefont
  {Dias}}, \bibinfo {author} {\bibfnamefont {G.~S.}\ \bibnamefont {Hartnett}},
  \bibinfo {author} {\bibfnamefont {M.}~\bibnamefont {Middleton}}, \bibinfo
  {author} {\bibfnamefont {P.}~\bibnamefont {Pani}}, \ and\ \bibinfo {author}
  {\bibfnamefont {J.~E.}\ \bibnamefont {Santos}},\ }\href {\doibase
  10.1088/1475-7516/2018/03/043} {\bibfield  {journal} {\bibinfo  {journal}
  {JCAP}\ }\textbf {\bibinfo {volume} {1803}},\ \bibinfo {pages} {043}
  (\bibinfo {year} {2018})},\ \Eprint {http://arxiv.org/abs/1801.01420}
  {arXiv:1801.01420 [gr-qc]} \BibitemShut {NoStop}%
\bibitem [{\citenamefont {Agrawal}\ \emph
  {et~al.}(2018{\natexlab{a}})\citenamefont {Agrawal}, \citenamefont
  {Marques-Tavares},\ and\ \citenamefont {Xue}}]{Agrawal:2017eqm}%
  \BibitemOpen
  \bibfield  {author} {\bibinfo {author} {\bibfnamefont {P.}~\bibnamefont
  {Agrawal}}, \bibinfo {author} {\bibfnamefont {G.}~\bibnamefont
  {Marques-Tavares}}, \ and\ \bibinfo {author} {\bibfnamefont {W.}~\bibnamefont
  {Xue}},\ }\href {\doibase 10.1007/JHEP03(2018)049} {\bibfield  {journal}
  {\bibinfo  {journal} {JHEP}\ }\textbf {\bibinfo {volume} {03}},\ \bibinfo
  {pages} {049} (\bibinfo {year} {2018}{\natexlab{a}})},\ \Eprint
  {http://arxiv.org/abs/1708.05008} {arXiv:1708.05008 [hep-ph]} \BibitemShut
  {NoStop}%
\bibitem [{\citenamefont {Anber}\ and\ \citenamefont
  {Sorbo}(2010)}]{Anber:2009ua}%
  \BibitemOpen
  \bibfield  {author} {\bibinfo {author} {\bibfnamefont {M.~M.}\ \bibnamefont
  {Anber}}\ and\ \bibinfo {author} {\bibfnamefont {L.}~\bibnamefont {Sorbo}},\
  }\href {\doibase 10.1103/PhysRevD.81.043534} {\bibfield  {journal} {\bibinfo
  {journal} {Phys. Rev.}\ }\textbf {\bibinfo {volume} {D81}},\ \bibinfo {pages}
  {043534} (\bibinfo {year} {2010})},\ \Eprint {http://arxiv.org/abs/0908.4089}
  {arXiv:0908.4089 [hep-th]} \BibitemShut {NoStop}%
\bibitem [{\citenamefont {Barnaby}\ \emph {et~al.}(2012)\citenamefont
  {Barnaby}, \citenamefont {Moxon}, \citenamefont {Namba}, \citenamefont
  {Peloso}, \citenamefont {Shiu},\ and\ \citenamefont {Zhou}}]{Barnaby:2012xt}%
  \BibitemOpen
  \bibfield  {author} {\bibinfo {author} {\bibfnamefont {N.}~\bibnamefont
  {Barnaby}}, \bibinfo {author} {\bibfnamefont {J.}~\bibnamefont {Moxon}},
  \bibinfo {author} {\bibfnamefont {R.}~\bibnamefont {Namba}}, \bibinfo
  {author} {\bibfnamefont {M.}~\bibnamefont {Peloso}}, \bibinfo {author}
  {\bibfnamefont {G.}~\bibnamefont {Shiu}}, \ and\ \bibinfo {author}
  {\bibfnamefont {P.}~\bibnamefont {Zhou}},\ }\href {\doibase
  10.1103/PhysRevD.86.103508} {\bibfield  {journal} {\bibinfo  {journal} {Phys.
  Rev.}\ }\textbf {\bibinfo {volume} {D86}},\ \bibinfo {pages} {103508}
  (\bibinfo {year} {2012})},\ \Eprint {http://arxiv.org/abs/1206.6117}
  {arXiv:1206.6117 [astro-ph.CO]} \BibitemShut {NoStop}%
\bibitem [{\citenamefont {Anber}\ and\ \citenamefont
  {Sorbo}(2012)}]{Anber:2012du}%
  \BibitemOpen
  \bibfield  {author} {\bibinfo {author} {\bibfnamefont {M.~M.}\ \bibnamefont
  {Anber}}\ and\ \bibinfo {author} {\bibfnamefont {L.}~\bibnamefont {Sorbo}},\
  }\href {\doibase 10.1103/PhysRevD.85.123537} {\bibfield  {journal} {\bibinfo
  {journal} {Phys. Rev.}\ }\textbf {\bibinfo {volume} {D85}},\ \bibinfo {pages}
  {123537} (\bibinfo {year} {2012})},\ \Eprint {http://arxiv.org/abs/1203.5849}
  {arXiv:1203.5849 [astro-ph.CO]} \BibitemShut {NoStop}%
\bibitem [{\citenamefont {Domcke}\ \emph {et~al.}(2016)\citenamefont {Domcke},
  \citenamefont {Pieroni},\ and\ \citenamefont {Binétruy}}]{Domcke:2016bkh}%
  \BibitemOpen
  \bibfield  {author} {\bibinfo {author} {\bibfnamefont {V.}~\bibnamefont
  {Domcke}}, \bibinfo {author} {\bibfnamefont {M.}~\bibnamefont {Pieroni}}, \
  and\ \bibinfo {author} {\bibfnamefont {P.}~\bibnamefont {Binétruy}},\ }\href
  {\doibase 10.1088/1475-7516/2016/06/031} {\bibfield  {journal} {\bibinfo
  {journal} {JCAP}\ }\textbf {\bibinfo {volume} {1606}},\ \bibinfo {pages}
  {031} (\bibinfo {year} {2016})},\ \Eprint {http://arxiv.org/abs/1603.01287}
  {arXiv:1603.01287 [astro-ph.CO]} \BibitemShut {NoStop}%
\bibitem [{\citenamefont {Garretson}\ \emph {et~al.}(1992)\citenamefont
  {Garretson}, \citenamefont {Field},\ and\ \citenamefont
  {Carroll}}]{Garretson:1992vt}%
  \BibitemOpen
  \bibfield  {author} {\bibinfo {author} {\bibfnamefont {W.~D.}\ \bibnamefont
  {Garretson}}, \bibinfo {author} {\bibfnamefont {G.~B.}\ \bibnamefont
  {Field}}, \ and\ \bibinfo {author} {\bibfnamefont {S.~M.}\ \bibnamefont
  {Carroll}},\ }\href {\doibase 10.1103/PhysRevD.46.5346} {\bibfield  {journal}
  {\bibinfo  {journal} {Phys. Rev.}\ }\textbf {\bibinfo {volume} {D46}},\
  \bibinfo {pages} {5346} (\bibinfo {year} {1992})},\ \Eprint
  {http://arxiv.org/abs/hep-ph/9209238} {arXiv:hep-ph/9209238 [hep-ph]}
  \BibitemShut {NoStop}%
\bibitem [{\citenamefont {Ratra}(1992)}]{Ratra:1991bn}%
  \BibitemOpen
  \bibfield  {author} {\bibinfo {author} {\bibfnamefont {B.}~\bibnamefont
  {Ratra}},\ }\href {\doibase 10.1086/186384} {\bibfield  {journal} {\bibinfo
  {journal} {Astrophys. J.}\ }\textbf {\bibinfo {volume} {391}},\ \bibinfo
  {pages} {L1} (\bibinfo {year} {1992})}\BibitemShut {NoStop}%
\bibitem [{\citenamefont {Anber}\ and\ \citenamefont
  {Sorbo}(2006)}]{Anber:2006xt}%
  \BibitemOpen
  \bibfield  {author} {\bibinfo {author} {\bibfnamefont {M.~M.}\ \bibnamefont
  {Anber}}\ and\ \bibinfo {author} {\bibfnamefont {L.}~\bibnamefont {Sorbo}},\
  }\href {\doibase 10.1088/1475-7516/2006/10/018} {\bibfield  {journal}
  {\bibinfo  {journal} {JCAP}\ }\textbf {\bibinfo {volume} {0610}},\ \bibinfo
  {pages} {018} (\bibinfo {year} {2006})},\ \Eprint
  {http://arxiv.org/abs/astro-ph/0606534} {arXiv:astro-ph/0606534 [astro-ph]}
  \BibitemShut {NoStop}%
\bibitem [{\citenamefont {Choi}\ \emph {et~al.}(2018)\citenamefont {Choi},
  \citenamefont {Kim},\ and\ \citenamefont {Sekiguchi}}]{Choi:2018dqr}%
  \BibitemOpen
  \bibfield  {author} {\bibinfo {author} {\bibfnamefont {K.}~\bibnamefont
  {Choi}}, \bibinfo {author} {\bibfnamefont {H.}~\bibnamefont {Kim}}, \ and\
  \bibinfo {author} {\bibfnamefont {T.}~\bibnamefont {Sekiguchi}},\ }\href
  {\doibase 10.1103/PhysRevLett.121.031102} {\bibfield  {journal} {\bibinfo
  {journal} {Phys. Rev. Lett.}\ }\textbf {\bibinfo {volume} {121}},\ \bibinfo
  {pages} {031102} (\bibinfo {year} {2018})},\ \Eprint
  {http://arxiv.org/abs/1802.07269} {arXiv:1802.07269 [hep-ph]} \BibitemShut
  {NoStop}%
\bibitem [{\citenamefont {Fujita}\ \emph {et~al.}(2015)\citenamefont {Fujita},
  \citenamefont {Namba}, \citenamefont {Tada}, \citenamefont {Takeda},\ and\
  \citenamefont {Tashiro}}]{Fujita:2015iga}%
  \BibitemOpen
  \bibfield  {author} {\bibinfo {author} {\bibfnamefont {T.}~\bibnamefont
  {Fujita}}, \bibinfo {author} {\bibfnamefont {R.}~\bibnamefont {Namba}},
  \bibinfo {author} {\bibfnamefont {Y.}~\bibnamefont {Tada}}, \bibinfo {author}
  {\bibfnamefont {N.}~\bibnamefont {Takeda}}, \ and\ \bibinfo {author}
  {\bibfnamefont {H.}~\bibnamefont {Tashiro}},\ }\href {\doibase
  10.1088/1475-7516/2015/05/054} {\bibfield  {journal} {\bibinfo  {journal}
  {JCAP}\ }\textbf {\bibinfo {volume} {1505}},\ \bibinfo {pages} {054}
  (\bibinfo {year} {2015})},\ \Eprint {http://arxiv.org/abs/1503.05802}
  {arXiv:1503.05802 [astro-ph.CO]} \BibitemShut {NoStop}%
\bibitem [{\citenamefont {Adshead}\ \emph {et~al.}(2016)\citenamefont
  {Adshead}, \citenamefont {Giblin}, \citenamefont {Scully},\ and\
  \citenamefont {Sfakianakis}}]{Adshead:2016iae}%
  \BibitemOpen
  \bibfield  {author} {\bibinfo {author} {\bibfnamefont {P.}~\bibnamefont
  {Adshead}}, \bibinfo {author} {\bibfnamefont {J.~T.}\ \bibnamefont {Giblin}},
  \bibinfo {author} {\bibfnamefont {T.~R.}\ \bibnamefont {Scully}}, \ and\
  \bibinfo {author} {\bibfnamefont {E.~I.}\ \bibnamefont {Sfakianakis}},\
  }\href {\doibase 10.1088/1475-7516/2016/10/039} {\bibfield  {journal}
  {\bibinfo  {journal} {JCAP}\ }\textbf {\bibinfo {volume} {1610}},\ \bibinfo
  {pages} {039} (\bibinfo {year} {2016})},\ \Eprint
  {http://arxiv.org/abs/1606.08474} {arXiv:1606.08474 [astro-ph.CO]}
  \BibitemShut {NoStop}%
\bibitem [{\citenamefont {Kitajima}\ \emph
  {et~al.}(2018{\natexlab{a}})\citenamefont {Kitajima}, \citenamefont
  {Sekiguchi},\ and\ \citenamefont {Takahashi}}]{Kitajima:2017peg}%
  \BibitemOpen
  \bibfield  {author} {\bibinfo {author} {\bibfnamefont {N.}~\bibnamefont
  {Kitajima}}, \bibinfo {author} {\bibfnamefont {T.}~\bibnamefont {Sekiguchi}},
  \ and\ \bibinfo {author} {\bibfnamefont {F.}~\bibnamefont {Takahashi}},\
  }\href {\doibase 10.1016/j.physletb.2018.04.024} {\bibfield  {journal}
  {\bibinfo  {journal} {Phys. Lett.}\ }\textbf {\bibinfo {volume} {B781}},\
  \bibinfo {pages} {684} (\bibinfo {year} {2018}{\natexlab{a}})},\ \Eprint
  {http://arxiv.org/abs/1711.06590} {arXiv:1711.06590 [hep-ph]} \BibitemShut
  {NoStop}%
\bibitem [{\citenamefont {Dror}\ \emph {et~al.}(2018)\citenamefont {Dror},
  \citenamefont {Harigaya},\ and\ \citenamefont {Narayan}}]{Dror:2018pdh}%
  \BibitemOpen
  \bibfield  {author} {\bibinfo {author} {\bibfnamefont {J.~A.}\ \bibnamefont
  {Dror}}, \bibinfo {author} {\bibfnamefont {K.}~\bibnamefont {Harigaya}}, \
  and\ \bibinfo {author} {\bibfnamefont {V.}~\bibnamefont {Narayan}},\
  }\href@noop {} {\  (\bibinfo {year} {2018})},\ \Eprint
  {http://arxiv.org/abs/1810.07195} {arXiv:1810.07195 [hep-ph]} \BibitemShut
  {NoStop}%
\bibitem [{\citenamefont {Co}\ \emph {et~al.}(2018)\citenamefont {Co},
  \citenamefont {Pierce}, \citenamefont {Zhang},\ and\ \citenamefont
  {Zhao}}]{Co:2018lka}%
  \BibitemOpen
  \bibfield  {author} {\bibinfo {author} {\bibfnamefont {R.~T.}\ \bibnamefont
  {Co}}, \bibinfo {author} {\bibfnamefont {A.}~\bibnamefont {Pierce}}, \bibinfo
  {author} {\bibfnamefont {Z.}~\bibnamefont {Zhang}}, \ and\ \bibinfo {author}
  {\bibfnamefont {Y.}~\bibnamefont {Zhao}},\ }\href@noop {} {\  (\bibinfo
  {year} {2018})},\ \Eprint {http://arxiv.org/abs/1810.07196} {arXiv:1810.07196
  [hep-ph]} \BibitemShut {NoStop}%
\bibitem [{\citenamefont {Bastero-Gil}\ \emph {et~al.}(2018)\citenamefont
  {Bastero-Gil}, \citenamefont {Santiago}, \citenamefont {Ubaldi},\ and\
  \citenamefont {Vega-Morales}}]{Bastero-Gil:2018uel}%
  \BibitemOpen
  \bibfield  {author} {\bibinfo {author} {\bibfnamefont {M.}~\bibnamefont
  {Bastero-Gil}}, \bibinfo {author} {\bibfnamefont {J.}~\bibnamefont
  {Santiago}}, \bibinfo {author} {\bibfnamefont {L.}~\bibnamefont {Ubaldi}}, \
  and\ \bibinfo {author} {\bibfnamefont {R.}~\bibnamefont {Vega-Morales}},\
  }\href@noop {} {\  (\bibinfo {year} {2018})},\ \Eprint
  {http://arxiv.org/abs/1810.07208} {arXiv:1810.07208 [hep-ph]} \BibitemShut
  {NoStop}%
\bibitem [{\citenamefont {Agrawal}\ \emph
  {et~al.}(2018{\natexlab{b}})\citenamefont {Agrawal}, \citenamefont
  {Kitajima}, \citenamefont {Reece}, \citenamefont {Sekiguchi},\ and\
  \citenamefont {Takahashi}}]{Agrawal:2018vin}%
  \BibitemOpen
  \bibfield  {author} {\bibinfo {author} {\bibfnamefont {P.}~\bibnamefont
  {Agrawal}}, \bibinfo {author} {\bibfnamefont {N.}~\bibnamefont {Kitajima}},
  \bibinfo {author} {\bibfnamefont {M.}~\bibnamefont {Reece}}, \bibinfo
  {author} {\bibfnamefont {T.}~\bibnamefont {Sekiguchi}}, \ and\ \bibinfo
  {author} {\bibfnamefont {F.}~\bibnamefont {Takahashi}},\ }\href@noop {} {\
  (\bibinfo {year} {2018}{\natexlab{b}})},\ \Eprint
  {http://arxiv.org/abs/1810.07188} {arXiv:1810.07188 [hep-ph]} \BibitemShut
  {NoStop}%
\bibitem [{\citenamefont {Hook}\ and\ \citenamefont
  {Marques-Tavares}(2016)}]{Hook:2016mqo}%
  \BibitemOpen
  \bibfield  {author} {\bibinfo {author} {\bibfnamefont {A.}~\bibnamefont
  {Hook}}\ and\ \bibinfo {author} {\bibfnamefont {G.}~\bibnamefont
  {Marques-Tavares}},\ }\href {\doibase 10.1007/JHEP12(2016)101} {\bibfield
  {journal} {\bibinfo  {journal} {JHEP}\ }\textbf {\bibinfo {volume} {12}},\
  \bibinfo {pages} {101} (\bibinfo {year} {2016})},\ \Eprint
  {http://arxiv.org/abs/1607.01786} {arXiv:1607.01786 [hep-ph]} \BibitemShut
  {NoStop}%
\bibitem [{\citenamefont {Fonseca}\ \emph {et~al.}(2018)\citenamefont
  {Fonseca}, \citenamefont {Morgante},\ and\ \citenamefont
  {Servant}}]{Fonseca:2018xzp}%
  \BibitemOpen
  \bibfield  {author} {\bibinfo {author} {\bibfnamefont {N.}~\bibnamefont
  {Fonseca}}, \bibinfo {author} {\bibfnamefont {E.}~\bibnamefont {Morgante}}, \
  and\ \bibinfo {author} {\bibfnamefont {G.}~\bibnamefont {Servant}},\
  }\href@noop {} {\  (\bibinfo {year} {2018})},\ \Eprint
  {http://arxiv.org/abs/1805.04543} {arXiv:1805.04543 [hep-ph]} \BibitemShut
  {NoStop}%
\bibitem [{\citenamefont {Soda}\ and\ \citenamefont
  {Urakawa}(2018)}]{Soda:2017dsu}%
  \BibitemOpen
  \bibfield  {author} {\bibinfo {author} {\bibfnamefont {J.}~\bibnamefont
  {Soda}}\ and\ \bibinfo {author} {\bibfnamefont {Y.}~\bibnamefont {Urakawa}},\
  }\href {\doibase 10.1140/epjc/s10052-018-6246-6} {\bibfield  {journal}
  {\bibinfo  {journal} {Eur. Phys. J.}\ }\textbf {\bibinfo {volume} {C78}},\
  \bibinfo {pages} {779} (\bibinfo {year} {2018})},\ \Eprint
  {http://arxiv.org/abs/1710.00305} {arXiv:1710.00305 [astro-ph.CO]}
  \BibitemShut {NoStop}%
\bibitem [{\citenamefont {Kitajima}\ \emph
  {et~al.}(2018{\natexlab{b}})\citenamefont {Kitajima}, \citenamefont {Soda},\
  and\ \citenamefont {Urakawa}}]{Kitajima:2018zco}%
  \BibitemOpen
  \bibfield  {author} {\bibinfo {author} {\bibfnamefont {N.}~\bibnamefont
  {Kitajima}}, \bibinfo {author} {\bibfnamefont {J.}~\bibnamefont {Soda}}, \
  and\ \bibinfo {author} {\bibfnamefont {Y.}~\bibnamefont {Urakawa}},\ }\href
  {\doibase 10.1088/1475-7516/2018/10/008} {\bibfield  {journal} {\bibinfo
  {journal} {JCAP}\ }\textbf {\bibinfo {volume} {1810}},\ \bibinfo {pages}
  {008} (\bibinfo {year} {2018}{\natexlab{b}})},\ \Eprint
  {http://arxiv.org/abs/1807.07037} {arXiv:1807.07037 [astro-ph.CO]}
  \BibitemShut {NoStop}%
\bibitem [{\citenamefont {Kim}\ \emph {et~al.}(2005)\citenamefont {Kim},
  \citenamefont {Nilles},\ and\ \citenamefont {Peloso}}]{Kim:2004rp}%
  \BibitemOpen
  \bibfield  {author} {\bibinfo {author} {\bibfnamefont {J.~E.}\ \bibnamefont
  {Kim}}, \bibinfo {author} {\bibfnamefont {H.~P.}\ \bibnamefont {Nilles}}, \
  and\ \bibinfo {author} {\bibfnamefont {M.}~\bibnamefont {Peloso}},\ }\href
  {\doibase 10.1088/1475-7516/2005/01/005} {\bibfield  {journal} {\bibinfo
  {journal} {JCAP}\ }\textbf {\bibinfo {volume} {0501}},\ \bibinfo {pages}
  {005} (\bibinfo {year} {2005})},\ \Eprint
  {http://arxiv.org/abs/hep-ph/0409138} {arXiv:hep-ph/0409138 [hep-ph]}
  \BibitemShut {NoStop}%
\bibitem [{\citenamefont {Agrawal}\ \emph
  {et~al.}(2018{\natexlab{c}})\citenamefont {Agrawal}, \citenamefont {Fan},\
  and\ \citenamefont {Reece}}]{Agrawal:2018mkd}%
  \BibitemOpen
  \bibfield  {author} {\bibinfo {author} {\bibfnamefont {P.}~\bibnamefont
  {Agrawal}}, \bibinfo {author} {\bibfnamefont {J.}~\bibnamefont {Fan}}, \ and\
  \bibinfo {author} {\bibfnamefont {M.}~\bibnamefont {Reece}},\ }\href@noop {}
  {\  (\bibinfo {year} {2018}{\natexlab{c}})},\ \Eprint
  {http://arxiv.org/abs/1806.09621} {arXiv:1806.09621 [hep-th]} \BibitemShut
  {NoStop}%
\bibitem [{\citenamefont {Sorbo}(2011)}]{Sorbo:2011rz}%
  \BibitemOpen
  \bibfield  {author} {\bibinfo {author} {\bibfnamefont {L.}~\bibnamefont
  {Sorbo}},\ }\href {\doibase 10.1088/1475-7516/2011/06/003} {\bibfield
  {journal} {\bibinfo  {journal} {JCAP}\ }\textbf {\bibinfo {volume} {1106}},\
  \bibinfo {pages} {003} (\bibinfo {year} {2011})},\ \Eprint
  {http://arxiv.org/abs/1101.1525} {arXiv:1101.1525 [astro-ph.CO]} \BibitemShut
  {NoStop}%
\bibitem [{\citenamefont {Giblin}\ and\ \citenamefont
  {Thrane}(2014)}]{Giblin:2014gra}%
  \BibitemOpen
  \bibfield  {author} {\bibinfo {author} {\bibfnamefont {J.~T.}\ \bibnamefont
  {Giblin}}\ and\ \bibinfo {author} {\bibfnamefont {E.}~\bibnamefont
  {Thrane}},\ }\href {\doibase 10.1103/PhysRevD.90.107502} {\bibfield
  {journal} {\bibinfo  {journal} {Phys. Rev.}\ }\textbf {\bibinfo {volume}
  {D90}},\ \bibinfo {pages} {107502} (\bibinfo {year} {2014})},\ \Eprint
  {http://arxiv.org/abs/1410.4779} {arXiv:1410.4779 [gr-qc]} \BibitemShut
  {NoStop}%
\bibitem [{\citenamefont {Buchmüller}\ \emph {et~al.}(2013)\citenamefont
  {Buchmüller}, \citenamefont {Domcke}, \citenamefont {Kamada},\ and\
  \citenamefont {Schmitz}}]{Buchmuller:2013lra}%
  \BibitemOpen
  \bibfield  {author} {\bibinfo {author} {\bibfnamefont {W.}~\bibnamefont
  {Buchmüller}}, \bibinfo {author} {\bibfnamefont {V.}~\bibnamefont {Domcke}},
  \bibinfo {author} {\bibfnamefont {K.}~\bibnamefont {Kamada}}, \ and\ \bibinfo
  {author} {\bibfnamefont {K.}~\bibnamefont {Schmitz}},\ }\href {\doibase
  10.1088/1475-7516/2013/10/003} {\bibfield  {journal} {\bibinfo  {journal}
  {JCAP}\ }\textbf {\bibinfo {volume} {1310}},\ \bibinfo {pages} {003}
  (\bibinfo {year} {2013})},\ \Eprint {http://arxiv.org/abs/1305.3392}
  {arXiv:1305.3392 [hep-ph]} \BibitemShut {NoStop}%
\bibitem [{\citenamefont {Tanabashi}\ \emph {et~al.}(2018)\citenamefont
  {Tanabashi} \emph {et~al.}}]{Tanabashi:2018oca}%
  \BibitemOpen
  \bibfield  {author} {\bibinfo {author} {\bibfnamefont {M.}~\bibnamefont
  {Tanabashi}} \emph {et~al.} (\bibinfo {collaboration} {Particle Data
  Group}),\ }\href {\doibase 10.1103/PhysRevD.98.030001} {\bibfield  {journal}
  {\bibinfo  {journal} {Phys. Rev.}\ }\textbf {\bibinfo {volume} {D98}},\
  \bibinfo {pages} {030001} (\bibinfo {year} {2018})}\BibitemShut {NoStop}%
\bibitem [{\citenamefont {Husdal}(2016)}]{Husdal:2016haj}%
  \BibitemOpen
  \bibfield  {author} {\bibinfo {author} {\bibfnamefont {L.}~\bibnamefont
  {Husdal}},\ }\href {\doibase 10.3390/galaxies4040078} {\bibfield  {journal}
  {\bibinfo  {journal} {Galaxies}\ }\textbf {\bibinfo {volume} {4}},\ \bibinfo
  {pages} {78} (\bibinfo {year} {2016})},\ \Eprint
  {http://arxiv.org/abs/1609.04979} {arXiv:1609.04979 [astro-ph.CO]}
  \BibitemShut {NoStop}%
\bibitem [{\citenamefont {Aghanim}\ \emph {et~al.}(2018)\citenamefont {Aghanim}
  \emph {et~al.}}]{Aghanim:2018eyx}%
  \BibitemOpen
  \bibfield  {author} {\bibinfo {author} {\bibfnamefont {N.}~\bibnamefont
  {Aghanim}} \emph {et~al.} (\bibinfo {collaboration} {Planck}),\ }\href@noop
  {} {\  (\bibinfo {year} {2018})},\ \Eprint {http://arxiv.org/abs/1807.06209}
  {arXiv:1807.06209 [astro-ph.CO]} \BibitemShut {NoStop}%
\bibitem [{\citenamefont {Abazajian}\ \emph {et~al.}(2016)\citenamefont
  {Abazajian} \emph {et~al.}}]{Abazajian:2016yjj}%
  \BibitemOpen
  \bibfield  {author} {\bibinfo {author} {\bibfnamefont {K.~N.}\ \bibnamefont
  {Abazajian}} \emph {et~al.} (\bibinfo {collaboration} {CMB-S4}),\ }\href@noop
  {} {\  (\bibinfo {year} {2016})},\ \Eprint {http://arxiv.org/abs/1610.02743}
  {arXiv:1610.02743 [astro-ph.CO]} \BibitemShut {NoStop}%
\bibitem [{\citenamefont {Kofman}\ \emph {et~al.}(1997)\citenamefont {Kofman},
  \citenamefont {Linde},\ and\ \citenamefont {Starobinsky}}]{Kofman:1997yn}%
  \BibitemOpen
  \bibfield  {author} {\bibinfo {author} {\bibfnamefont {L.}~\bibnamefont
  {Kofman}}, \bibinfo {author} {\bibfnamefont {A.~D.}\ \bibnamefont {Linde}}, \
  and\ \bibinfo {author} {\bibfnamefont {A.~A.}\ \bibnamefont {Starobinsky}},\
  }\href {\doibase 10.1103/PhysRevD.56.3258} {\bibfield  {journal} {\bibinfo
  {journal} {Phys. Rev.}\ }\textbf {\bibinfo {volume} {D56}},\ \bibinfo {pages}
  {3258} (\bibinfo {year} {1997})},\ \Eprint
  {http://arxiv.org/abs/hep-ph/9704452} {arXiv:hep-ph/9704452 [hep-ph]}
  \BibitemShut {NoStop}%
\bibitem [{\citenamefont {Dufaux}\ \emph {et~al.}(2006)\citenamefont {Dufaux},
  \citenamefont {Felder}, \citenamefont {Kofman}, \citenamefont {Peloso},\ and\
  \citenamefont {Podolsky}}]{Dufaux:2006ee}%
  \BibitemOpen
  \bibfield  {author} {\bibinfo {author} {\bibfnamefont {J.~F.}\ \bibnamefont
  {Dufaux}}, \bibinfo {author} {\bibfnamefont {G.~N.}\ \bibnamefont {Felder}},
  \bibinfo {author} {\bibfnamefont {L.}~\bibnamefont {Kofman}}, \bibinfo
  {author} {\bibfnamefont {M.}~\bibnamefont {Peloso}}, \ and\ \bibinfo {author}
  {\bibfnamefont {D.}~\bibnamefont {Podolsky}},\ }\href {\doibase
  10.1088/1475-7516/2006/07/006} {\bibfield  {journal} {\bibinfo  {journal}
  {JCAP}\ }\textbf {\bibinfo {volume} {0607}},\ \bibinfo {pages} {006}
  (\bibinfo {year} {2006})},\ \Eprint {http://arxiv.org/abs/hep-ph/0602144}
  {arXiv:hep-ph/0602144 [hep-ph]} \BibitemShut {NoStop}%
\bibitem [{\citenamefont {Adshead}\ \emph {et~al.}(2015)\citenamefont
  {Adshead}, \citenamefont {Giblin}, \citenamefont {Scully},\ and\
  \citenamefont {Sfakianakis}}]{Adshead:2015pva}%
  \BibitemOpen
  \bibfield  {author} {\bibinfo {author} {\bibfnamefont {P.}~\bibnamefont
  {Adshead}}, \bibinfo {author} {\bibfnamefont {J.~T.}\ \bibnamefont {Giblin}},
  \bibinfo {author} {\bibfnamefont {T.~R.}\ \bibnamefont {Scully}}, \ and\
  \bibinfo {author} {\bibfnamefont {E.~I.}\ \bibnamefont {Sfakianakis}},\
  }\href {\doibase 10.1088/1475-7516/2015/12/034} {\bibfield  {journal}
  {\bibinfo  {journal} {JCAP}\ }\textbf {\bibinfo {volume} {1512}},\ \bibinfo
  {pages} {034} (\bibinfo {year} {2015})},\ \Eprint
  {http://arxiv.org/abs/1502.06506} {arXiv:1502.06506 [astro-ph.CO]}
  \BibitemShut {NoStop}%
\end{thebibliography}
\end{document}